\documentclass[%
 reprint,
superscriptaddress,
 amsmath,amssymb,
 aps,
]{revtex4-1}

\usepackage[dvipsnames]{xcolor}

\usepackage{comment}

\usepackage[redeflists]{IEEEtrantools}

\usepackage{graphicx}
\usepackage{dcolumn}
\usepackage{bm}

\begin{document}


\title{Quantitative prediction of sling events in turbulence at high Reynolds numbers}
\author{Tobias B\"atge}
\affiliation{Max-Planck-Institute for Dynamics and Self-Organisation,  (MPI DS), Am Faßberg 17, 37077 G\"ottingen, Germany} 
\author{Itzhak Fouxon}
\affiliation{Department of Computational Science and Engineering, Yonsei University, Seoul 03722, South Korea}

\author{Michael Wilczek}
\affiliation{Max-Planck-Institute for Dynamics and Self-Organisation,  (MPI DS), Am Faßberg 17, 37077 G\"ottingen, Germany} 
\affiliation{Theoretical Physics I, University of Bayreuth, Universitätsstr. 30, 95447 Bayreuth, Germany}

\date{\today}
\begin{abstract}

Collisional growth of droplets, such as occurring in warm clouds, is known to be significantly enhanced by turbulence. Whether particles collide depends on their flow history, in particular on their encounters with highly intermittent small-scale turbulent structures, which despite their rarity can dominate the overall collision rate. Intuitively, strong vortices may act as slings for inertial particles, leading to intersections where several streams of particles collide at large velocities. Here, we develop a quantitative criterion for sling events based on the velocity gradient history along particle paths. We demonstrate by combination of theory and simulations that the problem reduces to a one-dimensional localization problem as encountered in condensed matter physics. The reduction demonstrates that the creation of slings is completely controlled by the smallest negative eigenvalue of the velocity gradient tensor.
We use fully resolved turbulence simulations to confirm our predictions and study their Stokes and Reynolds number dependence. We also discuss extrapolations to the parameter range relevant in clouds, showing that sling events at high Reynolds numbers are significantly enhanced for small Stokes numbers.

\end{abstract}

\maketitle

In warm clouds, microscopic water droplets tend to have much larger collisional velocities in rare spatial regions which may result in a significant enhancement of the volume-averaged collision rate \cite{Devenish2012}. As the droplets grow to form rain, first by vapor condensation on initial nuclei, and then by collisions and coalescence, they pass through a range of sizes where cloud turbulence plays a significant role \cite{Shaw2003, Pruppacher2010, Devenish2012,Grabowski2013,Pumir2016, Khain2018}.
Due to the intermittent nature of small-scale turbulence, droplets frequently encounter regions of strong vorticity or strain in which droplet inertia shows an appreciable effect, introducing spatial clustering and high relative velocities \cite{Zaichik2009, pan2010,ireland2016,petersen2019, Bertens2021}. In particular, droplets are ejected from very strong vortices as off slings -- giving rise to the sling effect \cite{Falkovich2002}. They detach from the air flow and shower on nearby calmer regions \cite{Falkovich2002,Wilkinson2005, bec2010}, confirmed experimentally in \cite{Bewley2013}. This creates regions of very frequent collisions which may dominate the overall collision rate \cite{Vosskuhle2014}. 

Quantitative estimates of the sling effect in clouds, which would help to solve the currently existing order-of-magnitude discrepancies in predicting rain formation times, today are still elusive \cite{Bhatnagar2022}. 
 Since the largest Reynolds number $\text{Re}_{\lambda}$ attainable in today's numerical simulations is much smaller than $10^4$ that may occur in clouds \cite{Devenish2012, Risius2015}, a major open question is to develop quantitative criteria to predict particle collisions and to extrapolate them to Reynolds number ranges relevant in clouds. 
In this context, the frequency at which droplets encounter the sling effect along their trajectory is a key ingredient to the collision kernel \cite{Falkovich2002} and can have a significant contribution \cite{Vosskuhle2014}. Current work towards characterizing the flow conditions leading to sling events \cite{Meibohm2021, Meibohm2022} emphasizes the importance of quantifying this frequency.

In this Letter, we derive criterion for the occurrence of sling events and provide a prediction for the $\text{Re}_{\lambda}$ dependence of the frequency of sling events by combining theory, recent rigorous results on the high Reynolds number limit \cite{Fouxon2020}, and fully resolved turbulence simulations. Remarkably, we demonstrate that the problem of the occurrence of the sling events can be effectively reduced to a one-dimensional problem, which coincides with that of a $1d$ Anderson localization problem \cite{lifshits1988}. In contrast to the previously proposed reduction as a $1d$ model of $3d$ collisions \cite{Derevyanko2007}, we here demonstrate the actual reduction by identifying the random potential of the $1d$ Schr\"odinger equation as given by the minimal real eigenvalue of turbulent velocity gradient tensor. 

\begin{figure}
    \centering
    \includegraphics[width=\linewidth]{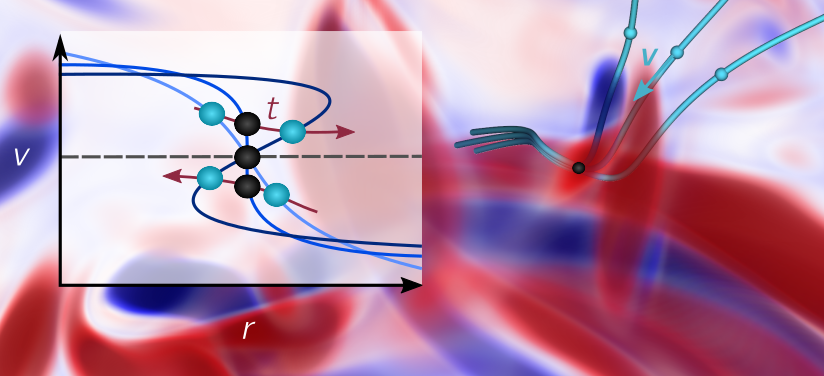}
    \caption{A volume rendering of the second invariant of the fluid flow velocity gradient tensor $Q = -\frac{1}{2}\mathrm{Tr}(\mathrm{A}^2)$ shows strain-dominated regions in red and vorticity-dominated regions in blue. The middle light blue trajectory originates from a simulation, whereas the two accompanying ones are obtained by linearizing the velocity field. A sling event is marked by a intersection of trajectories. This is analogous to a one-dimensional caustic as visualized in the inset.}
    \label{fig:1}
\end{figure}

We model droplets as inertial particles, which allows to separate the effects of turbulence from more complex settings including thermodynamics and hydrodynamic interactions \cite{Devenish2012,Saito2018, Li2020}. Already in this simple setting, the physics of particle collisions is quite rich. When the inertia of a particle is very small and its velocity relaxation to the ambient flow is fast, collision rates can be predicted in the framework of the Saffmann-Turner theory which neglects inertia. Particles move as tracers in a smooth flow and collide only because their size is small but finite \cite{Saffman1956}. Particle inertia introduces fundamentally new effects such as centrifugal forces. The particles can still be considered as tracers, however in an effective flow which is compressible and differs from the underlying incompressible turbulence \cite{Maxey1987, Balkovsky2001}. Compressibility leads to preferential concentration, i.e.~particles distribute over a multifractal attractor set in real space \cite{Falkovich2004}. Again collisions occur only due to finite particle size. However, for strong vortices the effective tracer description fails and point particles can collide due to the sling effect, i.e.~the intersection of trajectories of nearby particles as in Fig.~\ref{fig:1}. The sling effect is concentrated in rare spatio-temporal regions inside which the particles' flow is three-valued, in contrast to outside single-valued smooth flow.

In our approach, the coordinates $\bm x$ and velocity $\bm v$ of each particle obey \footnote{The equations apply for all droplets sizes of interest provided that $\tau$ is considered as effective time, see e.g. 
\cite{Fouxon2015}}
\begin{eqnarray}&& \!\!\!\!\!\!\!\!
\dot{\bm x}(t)\!=\!\bm v(t),\ \ \tau\dot{\bm v}(t)=\bm u(\bm x(t), t)+\bm v_g-\bm v(t). \label{eq:motion}
\end{eqnarray}
The equations describe relaxation of the particle velocity to the sum of the local turbulent velocity $\bm u(\bm x, t)$ and the gravitational settling velocity in still air $\bm v_g$ with the Stokes time $\tau$.
The system can be characterized by three dimensionless parameters: the Stokes number $\text{St}=\tau\sqrt{\epsilon/\nu}$ where $\epsilon$ is the mean energy dissipation rate and $\nu$ is the kinematic viscosity, the Froude number $\text{Fr}=\epsilon^{3/4}/(g\nu^{1/4})$, where $g$ denotes the gravitational acceleration, and the Taylor-microscale Reynolds number $\text{Re}_{\lambda} = \sqrt{15/(\nu\epsilon)} u^2$, where $u$ is the rms velocity component. All these parameters influence the collisions strongly ($\text{Re}_{\lambda}$ via the intermittency of turbulence), creating a complex problem.  

Eqs.~\eqref{eq:motion} generate a smooth flow in the six-dimensional single-particle phase space. However, after an initial transient, that flow is fully driven by the real space flow $\bm u(\bm x, t)$.
 Under certain conditions, like for example in the limit of negligible gravity and small but finite inertia \cite{Maxey1983}, one can then introduce a three-dimensional ``particle flow" $\bm v(\bm x, t)$ so that ${\bm v}\!=\!\bm v(\bm x(t), t)$ i.e.~the particle's velocity is uniquely defined by the position. If the flow exists, it has to obey
\begin{eqnarray}&& \!\!\!\!\!\!\!\!
\tau\left(\partial_t\bm v+ \bm v\cdot\nabla\bm v\right)=\bm u+\bm v_g-\bm v\,, \label{eq:pde}
\end{eqnarray}
to agree with Eqs.~\eqref{eq:motion}. If the evolution by this PDE generates a single-valued $\bm v(\bm x, t)$, the assumption is self-consistent. Conversely, if the solution becomes multi-valued at some time, which implies that some of the flow derivatives blow up at that time, then $\bm v(\bm x, t)$ does not exist in the blowup region. This reduces the problem to the study of generation of finite-time blowups of velocity gradients by Eq.~\eqref{eq:pde}. One can see from Eq.~\eqref{eq:pde} that $P_{ik}(t)\equiv \nabla_kv_i (\bm x(t), t)$ evolves according to
\begin{equation}
\tau\left(\dot {P}+P^2\right) = A-P,\label{eq:sigma}    
\end{equation}
where $A_{ik}(t)\equiv \nabla_ku_i (\bm x(t), t)$. Blowups happen if the history of gradients $A(t)$ along the particle's path $\bm x(t)$ produces $|P\tau|\gtrsim 1$. The $P^2$ term then starts to dominate in Eq.~\eqref{eq:sigma}, producing a finite-time singularity. The blowups form in regions which cause coexistence of three, or more, streams of particles (``folds") whose lifetime is of order $\tau$, see Fig.~\ref{fig:1} and \cite{Falkovich2002}.

\begin{figure}
    \centering
    \includegraphics{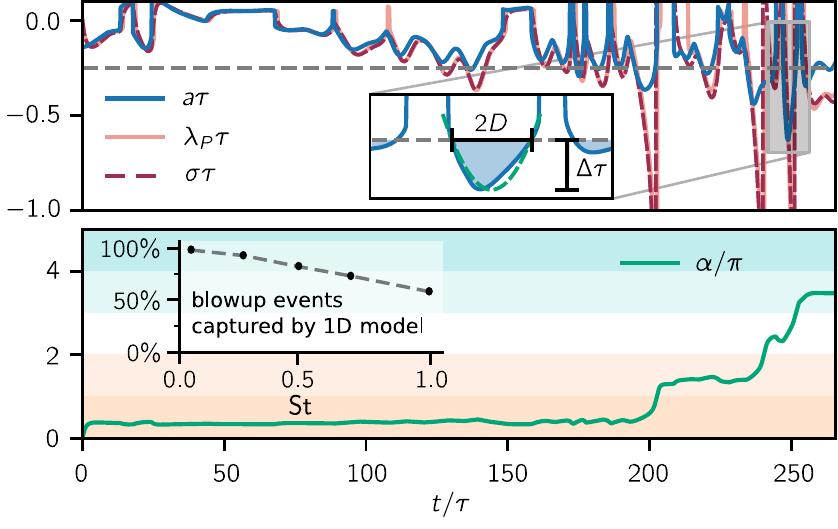}
    \caption{Sling events along an individual trajectory in a flow with $\text{Re}_\lambda=193$. The top panel shows the most negative real eigenvalue $a$ of the sampled velocity gradient, the most negative real eigenvalue $\lambda_P$ of the particle velocity gradient, which is obtained by solving the full matrix problem Eq.~\eqref{eq:sigma}, and its one-dimensional proxy $\sigma$, which is obtained by solving Eq.~\eqref{eq:sigmaf}. Sling events correspond to the divergence of $\sigma$ to negative infinity.
    The inset shows an excursion of $a$ below the critical value of $-1/(4\tau)$ (gray dashed line) with duration $2D$ and depth $\Delta$. Here, the green dashed line denotes the parabolic approximation discussed in the main text. The bottom panel shows the corresponding one-dimensional angle dynamics. Here, the sling events correspond to crossing multiples of $\pi$. For the given trajectory with $St=0.3$, the one-dimensional dynamics captures the occurrence of the sling events well. This is quantified in the inset, which shows the percentage of the sling events that are captured by the one-dimensional approximation \eqref{eq:sigmaf} compared to the full three-dimensional dynamics \eqref{eq:sigma} as a function of the Stokes number.}
    \label{fig:2}
\end{figure}
Crucially, the study of blowups in the matrix equation \eqref{eq:sigma} can be reduced to the study of blowups in a scalar equation. This essentially reflects the one-dimensional physics of the sling effect: faster particles moving in a certain direction pass slower particles in front of them \cite{Falkovich2002,Derevyanko2007, Meibohm2021, Meibohm2022}. However until now, the one-dimensionality has not been turned into a quantitative tool. To establish this, we first observe empricially that a rare blowup event consists roughly of two  stages.
Initially, $P(t)$ has a typical value with $|P\tau|\ll 1$ which approximately can be set to zero in considering growth to $|P\tau|\sim 1$. Then it starts to grow according to $P\tau=\int_0^t \exp(-(t-t')/\tau)A(t')dt'$ due to the trajectory encountering a rare large $A$, which is persistent on the scale of $\tau$. We take $A$ as quasi-constant since its significant oscillations would create ineffective growth of $P$ having much smaller probability. Accordingly, $P$ grows proportionally to $A$, and hence commutes with it. As a result, the system can be diagonalized and the eigenvalues follow the same equation as the matrix. Once it reaches a value $P\sim 1/\tau$, Eq.~\eqref{eq:sigma} reduces to $\dot {P}\approx -P^2$, i.e.~the evolution equation that generates finite time blowup of the most negative eigenvalue $\sigma(t)$ of $P$. Therefore 
\begin{eqnarray}&& \!\!\!\!\!\!\!\!
\tau\left(\dot {\sigma}+\sigma^2\right)\!=\!a-\sigma, \label{eq:sigmaf}
\end{eqnarray}
where $a$ is the most negative eigenvalue of $A$, provides a qualitatively valid description of the whole blowup process, see SI for detailed discussion. 

We can quantitatively check the validity of Eq.~\eqref{eq:sigmaf} with simulations by comparing the solutions of the full matrix equation \eqref{eq:sigma} to the one-dimensional approximation \eqref{eq:sigmaf}.
To this end, we obtained particle trajectories along with their full velocity gradient history from pseudo-spectral Navier-Stokes simulations covering a Reynolds number range up to $\text{Re}_{\lambda}\approx 500$ and taking $\text{Fr}\rightarrow\infty$, i.e.~neglecting gravity. We use a third-order Runge-Kutta time stepping scheme for the fields and a Heun scheme for the particle integration \cite{Lalescu2022}. In Fig.~\ref{fig:2} (top), we illustrate the direct correspondence between blowup events in the full 3D matrix dynamics and the one-dimensional approximation for a sample trajectory.

We detect blowups as crossings of integer multiples of $\pi$ by the finite angle variable $\cot \alpha\equiv  \sigma+(2\tau)^{-1}$, which avoids dealing with infinities, see SI. For the sample trajectory shown in Fig.~\ref{fig:2} (top), this is illustrated in Fig.~\ref{fig:2} (bottom). Based on this, for all Reynolds numbers under consideration, we find that  Eq.~\eqref{eq:sigmaf} captures the blowup event in about $95$ per cent of cases for small Stokes numbers up to $\text{St}=0.3$ whereas this rate decreases with increasing Stokes numbers, see inset in Fig.~\ref{fig:2} (bottom). 

Having established the validity of a one-dimensional description, we can derive a quantitative criterion to predict blowups. The angle $\alpha$ is well-known in $1d$ localization problems \cite{lifshits1988} as it is related to the logarithmic derivative of the wave function $\psi$, which obeys the
one-dimensional stationary Schr\"odinger equation
\begin{equation}
-\frac{d^2\psi}{dt^2}+\frac{a}{\tau}\psi=-\frac{\psi}{4\tau^2};\quad \frac{{\dot \psi}}{\psi}\equiv \sigma+\frac{1}{2\tau}=\cot \alpha\,. \label{eq:a}    
\end{equation}
Here $a/\tau$ is the random ``potential", $-(4\tau^2)^{-1}$ is the ``energy", and $t$ plays the role of the spatial coordinate. The blowups of $\sigma$ coincide with zeros of $\psi$. Most of the time the amplitude of the potential is much smaller than the one of the energy and constitutes a small perturbation. This type of motion is quasi-classical and can be described by \cite{Landau2013}
\begin{equation}
\psi(t) \sim \frac{\exp\left(\int^t p(s) ds\right)}{\sqrt{p(t)}},\ \
p(t)\equiv \sqrt{\frac{1}{4\tau^2}\!+\!\frac{a(t)}{\tau}}\,, \label{eq:qu}    
\end{equation}
where we keep only the exponentially growing solution, discarding the transients. This solution does not have zeros. One can see by inserting Eq.~\eqref{eq:qu} into Eq.~\eqref{eq:a} that its validity demands that $8\tau^2|\dot a|\ll |1+4\tau a|^{3/2}$. This condition is violated near the turning points of the potential, defined by $a=-1/(4\tau)$ where the right-hand side is small. The zeros of $\psi$ (blowups) occur in the classically allowed region where the energy $-(4\tau^2)^{-1}$ is larger than the potential $a/\tau$ and the wavefunction oscillates. In order to construct a globally valid solution that includes the classically allowed regions, we study rare excursions of $a$ to values smaller than $-1/(4\tau)$. These excursions are typically well-separated in time so we can concentrate on the effect of one excursion which is associated with a minimum of $a(t)$ which is smaller than $-1/(4\tau)$. To obtain an analytical criterion, we perform a quadratic expansion near the minimum which we set at $t=0$. This leads to the Weber equation \cite{Weber1869}
\begin{eqnarray} 
 -\frac{d^2\psi}{dt^2}+\frac{\Delta }{D^2}t^2\psi=\Delta \psi;\ \ \ \frac{a}{\tau}+\frac{1}{4\tau^2}\approx-\Delta+\frac{\Delta}{D^2}t^2 ,\label{eq:as}
\end{eqnarray}
where we defined $\Delta$ as the magnitude by which the minimum of $a(t)$ exceeds the threshold $-1/(4\tau)$ and $D$ as the duration of such excursions. The inset in Fig.~\ref{fig:2} shows such an excursion and the corresponding parabolic approximation. By solving the equation, we can construct a globally valid approximation for $\psi$, see the SI. The solution demonstrates that
the blowups of $\sigma$ happen only between the turning points in the quasi-classical region with $a<-1/(4\tau)$. The existence of the zero demands that
\begin{eqnarray} 
\sqrt{\Delta} D>1\,.
\label{eq:criterion}
\end{eqnarray}
\begin{figure}
    \centering
    \includegraphics[width=\linewidth]{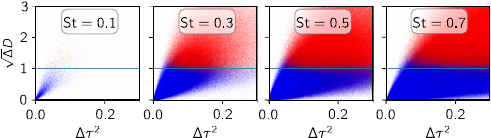}
    \caption{Scatter plot of depth and duration of excursions of the sampled most negative  real  eigenvalue  of  the  velocity  gradient for different Stokes numbers.   They  are  plotted in red if a  sling  event can be directly associated in the corresponding one-dimensional dynamics Eq.~\eqref{eq:sigmaf} and blue otherwise. The criterion \eqref{eq:criterion} (blue line) separates sling events from non-sling events. }
    \label{fig:3}
\end{figure}
Thus the product of magnitude of the excursion below $-1/(4\tau)$ and duration must be larger than $1$; to create a blowup, $|a(t)|$ must not only become very large but also act for a sufficient time. Furthermore, if the product is also smaller than $3$, $\psi$ has only one zero and accordingly the excursion produces only one blowup. However, if it is between $3$ and $5$ then two blowups occur in a row. We observed both one- and two-blowup events at the considered $\text{St}$. The probability of three-blowup events, that would happen if the product were higher than $5$, was found to be negligible for small $\text{St}$. 

To compare this criterion to DNS data, we determine the duration of the excursions and their depths from time series of the most negative eigenvalue $a$ of the velocity gradient and solve the one-dimensional dynamics \eqref{eq:sigmaf}. As Fig.~\ref{fig:3} shows, our criterion \eqref{eq:criterion} indeed separates well excursions that lead to at least one sling event from those that do not. Remarkably, it even performs well for Stokes numbers as high as $\text{St}=0.7$ where the above theory, which assumes a small probability of excursions, is not expected to work. Unfortunately, despite that Eq.~\eqref{eq:criterion} describes blowups of Eq.~\eqref{eq:sigmaf} at $\text{St}=0.7$ accurately, the correspondence between full matrix dynamics and one-dimensional approach is already weak, see inset in Fig.~\ref{fig:2}. Eq.~\eqref{eq:criterion} describes accurately about $80\%$ of blowups of Eq.~\eqref{eq:sigma} at $\text{St}=0.5$ which we propose as heuristic upper limit of the theory.

\begin{figure}
    \centering
    \includegraphics[width=\linewidth]{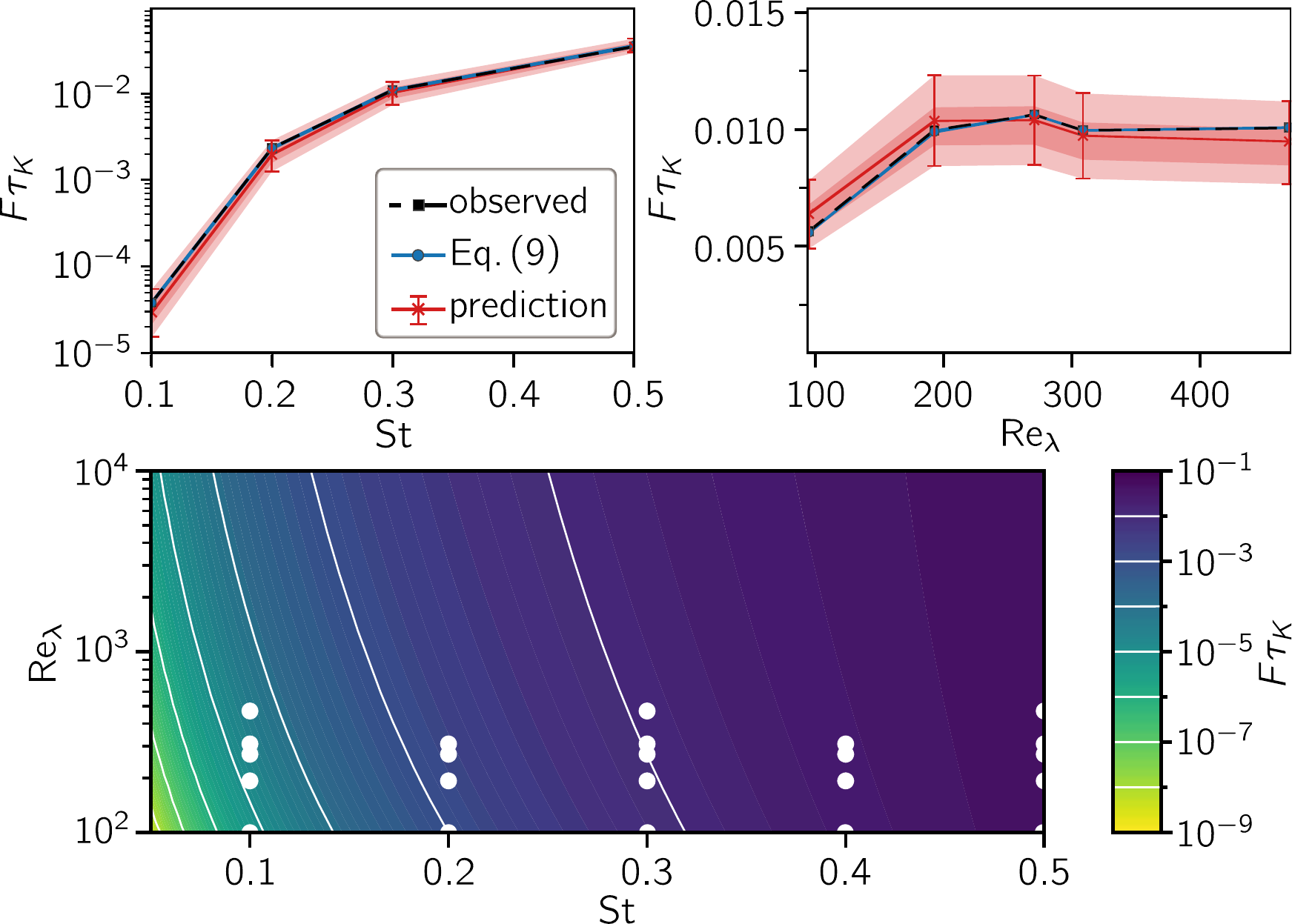}
    \caption{ Sling rate as a function of Stokes and Reynolds number.
    The black dashed line in the top panels corresponds to the sling rate obtained from DNS data using the one-dimensional description; the blue line is the prediction based on Eq.~\eqref{eq:sling_rate1}. The errorbars are 75\% and 95\% percentiles which are obtained by a bootstrap over the observed parameter ranges. (top left) Stokes number varies at a Reynolds number of $\text{Re}_\lambda=271$. (top right) Stokes number is set to $\text{St}=0.3$ and we vary the Reynolds number. (bottom) Extrapolated prediction of the sling rate as a function of $\text{St}$ and $\text{Re}_\lambda$. The white dots mark the DNS data sets used for the extrapolation.}
    \label{fig:4}
\end{figure}

Based on these findings, we can make statements about the frequency of sling events as a function of Reynolds number.
The frequency $F$ of the sling events can be obtained as:
\begin{equation}
    F=fP\left(\sqrt{\Delta}D>1\right)\, ,\label{eq:sling_rate1}
\end{equation}
i.e.~the frequency $f$ of excursions of $a(t)$ below $-1/(4\tau)$ times the conditional probability that the excursion obeys the criterion~\eqref{eq:criterion} given that it reaches below $-1/(4\tau)$. Using the observed $f$, $D$ and $\Delta$ along trajectories from our DNS, we demonstrate the validity of the above equation in Fig.~\ref{fig:4} as a function of Stokes and Reynolds number. Across the range of Stokes and Reynolds numbers investigated here, we find excellent quantitative agreement. 

To turn Eq.~\eqref{eq:sling_rate1} into a prediction for the rate of sling events $F$, which can also be extrapolated to higher Reynolds numbers, we need to characterize the frequency of excursions $f$, the probability of excursions leading to sling events, and their dependence on $\text{Re}_\lambda$ and $\text{St}$.
The excursion frequency is the fraction of time spent below the threshold value divided by the average duration of the individual excursions:
\begin{align}
f= \frac{1}{2\langle D\rangle}\int_{-\infty}^{-(4\tau)^{-1}}P(a)da\,,\label{eq:f}
\end{align}
where $P(a)$ is the PDF of $a(t)$. The integral above is the probability of excursions. We expect it to increase with the Reynolds number due to intermittency \cite{Frisch1995}. In contrast, $\langle D\rangle$, which is a typical duration of most probable excursions, 
should be approximately Reynolds number independent (but dependent on the Stokes number). For the average duration and its Reynolds independence, we make a phenomenological argument. The inverse of the threshold value  $-1/(4\tau)$ defines a timescale which is proportional to $\text{St}$ and independent of $\text{Re}_\lambda$. Assuming the characteristic timescale for an excursions behaves similarly, the mean duration should be proportional to the Stokes number, $\langle D\rangle\propto\text{St}\tau_K$. The proportionality constant is determined in the SI.

 To evaluate the integral, we note that
for large negative fluctuations of $a(t)$, $\nu a^2(t)$ should be approximately proportional to the local dissipation rate. Therefore we anticipate that the relevant left tail of $P(a)$ is directly related to the right tail of the PDF of the dissipation rate, which was observed to be log-normal in very good approximation \cite{Fouxon2020}.
 The log-normal approximation should therefore also work well for $P(a)$.  One can validate this by using large-deviation theory and our DNS results, where $\text{Re}_\lambda$ is the corresponding large parameter.
Due to the log-normal approximation for $P(a)$, the integral in Eq.~\eqref{eq:f} can be evaluated analytically and features an explicit dependence on $\text{Re}_\lambda$ and $\text{St}$, for more details we refer to the SI.
 
Finally, we turn to the probability $P(\sqrt{\Delta}D>1)$. For reasons similar to those for $\langle D\rangle$, we expect it to have only a weak Reynolds number dependence. This is confirmed by our DNS data.  To fix the Stokes number dependence, it is useful to observe that $P(\sqrt{\Delta}D>1)$ is well-defined for any $\text{St}$ and study its behavior at $0<\text{St}<\infty$. In the limit $\text{St}\rightarrow0$, this probability goes to zero as well, since the average duration tends to zero for $\text{St}\rightarrow0$, requiring infinitely deep excursions in this limit which are impossible. The limit $\text{St}\rightarrow\infty$ is a finite number which gives the probability that negative $a$ obeys Eq.~\eqref{eq:criterion} and is smaller than one. Since $P(\sqrt{\Delta}D>1)$ expected to to grow with $\text{St}$ monotonously, this suggests that it can be captured by an odd sigmoid function with two parameters controlling its amplitude and steepness. This is confirmed in the SI. \\
Above, we described how to characterize all the relevant terms in Eq.~\eqref{eq:sling_rate1}. The result is a quantitative description of the sling rate and its dependence on $\text{Re}_\lambda$ and $\text{St}$ (see SI for an explicit formula).

Equipped with this, we can fit our results as a function of the Reynolds and Stokes numbers. In Fig.~\ref{fig:4}, we show that this fit reproduces the observed sling rates in good approximation. Importantly, these fits allow an extrapolation of the results to higher Reynolds numbers as relevant in clouds.
The extrapolation of our prediction in Fig.~\ref{fig:4} shows that the effect of high Reynolds numbers becomes appreciable for lower Stokes numbers. For $\text{St}=0.1$ and $\text{Re}_\lambda=10^4$, our extrapolation predicts an increase of roughly an order of magnitude  $F\approx2\times10^{-4}\tau_K$ compared to the value for $\text{Re}_\lambda=271$. In contrast, for $\text{St}=0.3$ and $\text{Re}_\lambda=10^4$, we obtain $F\approx0.016\tau_K$, which is comparable to the results at moderate Reynolds numbers featured in Fig.~\ref{fig:4}. The stronger Reynolds dependence of the sling effect for small Stokes numbers might cause it to become a dominant mechanism of droplet growth in clouds also for small droplets. A precise prediction, however, about the collisions for a given sling rate requires future research about the timescale and spatial extent of sling events and their respective dependence on $\text{Re}_\lambda$. However one can make an estimate: using our prediction and an expression for the collision contribution of slings \cite{Falkovich2002}, we estimate in the SI that even for Stokes numbers $\text{St}=0.15$ or smaller the ratio of collisions caused by sling events changes more than an order of magnitude going from moderate Reynolds numbers to those in clouds.

In summary, we introduced an analytic approach to the sling frequency that can be extrapolated to large $\text{Re}_{\lambda}$, enabling predictions at high $\text{Re}_{\lambda}$ in clouds. 
We confirmed the theory with simulations of particle motion in Navier-Stokes turbulence. 
Future studies are to include gravity computationally and provide the full formula for the collision kernel of the cloud droplets theoretically. 

\section*{Acknowledgements}

This work was supported by the Fraunhofer -- Max-Planck Cooperation Program. We thank Eberhard Bodenschatz for helpful discussions. We thank Cristian C. Lalescu and B\'{e}renger Bramas for their support and development of the used TurTLE code. Computational resources from the Max Planck Computing and Data Facility and support by the Max Planck Society are gratefully acknowledged.

\bibliography{reference}

\begin{thebibliography}{41}%
\makeatletter
\providecommand \@ifxundefined [1]{%
 \@ifx{#1\undefined}
}%
\providecommand \@ifnum [1]{%
 \ifnum #1\expandafter \@firstoftwo
 \else \expandafter \@secondoftwo
 \fi
}%
\providecommand \@ifx [1]{%
 \ifx #1\expandafter \@firstoftwo
 \else \expandafter \@secondoftwo
 \fi
}%
\providecommand \natexlab [1]{#1}%
\providecommand \enquote  [1]{``#1''}%
\providecommand \bibnamefont  [1]{#1}%
\providecommand \bibfnamefont [1]{#1}%
\providecommand \citenamefont [1]{#1}%
\providecommand \href@noop [0]{\@secondoftwo}%
\providecommand \href [0]{\begingroup \@sanitize@url \@href}%
\providecommand \@href[1]{\@@startlink{#1}\@@href}%
\providecommand \@@href[1]{\endgroup#1\@@endlink}%
\providecommand \@sanitize@url [0]{\catcode `\\12\catcode `\$12\catcode
  `\&12\catcode `\#12\catcode `\^12\catcode `\_12\catcode `\%12\relax}%
\providecommand \@@startlink[1]{}%
\providecommand \@@endlink[0]{}%
\providecommand \url  [0]{\begingroup\@sanitize@url \@url }%
\providecommand \@url [1]{\endgroup\@href {#1}{\urlprefix }}%
\providecommand \urlprefix  [0]{URL }%
\providecommand \Eprint [0]{\href }%
\providecommand \doibase [0]{http://dx.doi.org/}%
\providecommand \selectlanguage [0]{\@gobble}%
\providecommand \bibinfo  [0]{\@secondoftwo}%
\providecommand \bibfield  [0]{\@secondoftwo}%
\providecommand \translation [1]{[#1]}%
\providecommand \BibitemOpen [0]{}%
\providecommand \bibitemStop [0]{}%
\providecommand \bibitemNoStop [0]{.\EOS\space}%
\providecommand \EOS [0]{\spacefactor3000\relax}%
\providecommand \BibitemShut  [1]{\csname bibitem#1\endcsname}%
\let\auto@bib@innerbib\@empty
\bibitem [{\citenamefont {Devenish}\ \emph {et~al.}(2012)\citenamefont
  {Devenish}, \citenamefont {Bartello}, \citenamefont {Brenguier},
  \citenamefont {Collins}, \citenamefont {Grabowski}, \citenamefont
  {IJzermans}, \citenamefont {Malinowski}, \citenamefont {Reeks}, \citenamefont
  {Vassilicos}, \citenamefont {Wang},\ and\ \citenamefont
  {Warhaft}}]{Devenish2012}%
  \BibitemOpen
  \bibfield  {author} {\bibinfo {author} {\bibfnamefont {B.~J.}\ \bibnamefont
  {Devenish}}, \bibinfo {author} {\bibfnamefont {P.}~\bibnamefont {Bartello}},
  \bibinfo {author} {\bibfnamefont {J.-L.}\ \bibnamefont {Brenguier}}, \bibinfo
  {author} {\bibfnamefont {L.~R.}\ \bibnamefont {Collins}}, \bibinfo {author}
  {\bibfnamefont {W.~W.}\ \bibnamefont {Grabowski}}, \bibinfo {author}
  {\bibfnamefont {R.~H.~A.}\ \bibnamefont {IJzermans}}, \bibinfo {author}
  {\bibfnamefont {S.~P.}\ \bibnamefont {Malinowski}}, \bibinfo {author}
  {\bibfnamefont {M.~W.}\ \bibnamefont {Reeks}}, \bibinfo {author}
  {\bibfnamefont {J.~C.}\ \bibnamefont {Vassilicos}}, \bibinfo {author}
  {\bibfnamefont {L.-P.}\ \bibnamefont {Wang}}, \ and\ \bibinfo {author}
  {\bibfnamefont {Z.}~\bibnamefont {Warhaft}},\ }\href {\doibase
  10.1002/qj.1897} {\bibfield  {journal} {\bibinfo  {journal} {Q. J. R.
  Meteorol. Soc.}\ }\textbf {\bibinfo {volume} {138}},\ \bibinfo {pages} {1401}
  (\bibinfo {year} {2012})}\BibitemShut {NoStop}%
\bibitem [{\citenamefont {Shaw}(2003)}]{Shaw2003}%
  \BibitemOpen
  \bibfield  {author} {\bibinfo {author} {\bibfnamefont {R.~A.}\ \bibnamefont
  {Shaw}},\ }\href {\doibase 10.1146/annurev.fluid.35.101101.161125} {\bibfield
   {journal} {\bibinfo  {journal} {Annu. Rev. Fluid Mech.}\ }\textbf {\bibinfo
  {volume} {35}},\ \bibinfo {pages} {183} (\bibinfo {year} {2003})}\BibitemShut
  {NoStop}%
\bibitem [{\citenamefont {Pruppacher}\ and\ \citenamefont
  {Klett}(2010)}]{Pruppacher2010}%
  \BibitemOpen
  \bibfield  {author} {\bibinfo {author} {\bibfnamefont {H.}~\bibnamefont
  {Pruppacher}}\ and\ \bibinfo {author} {\bibfnamefont {J.}~\bibnamefont
  {Klett}},\ }\href {\doibase 10.1007/978-0-306-48100-0} {\emph {\bibinfo
  {title} {{Microphysics of Clouds and Precipitation}}}},\ \bibinfo {series}
  {Atmospheric and Oceanographic Sciences Library}, Vol.~\bibinfo {volume}
  {18}\ (\bibinfo  {publisher} {Springer Netherlands},\ \bibinfo {address}
  {Dordrecht},\ \bibinfo {year} {2010})\BibitemShut {NoStop}%
\bibitem [{\citenamefont {Grabowski}\ and\ \citenamefont
  {Wang}(2013)}]{Grabowski2013}%
  \BibitemOpen
  \bibfield  {author} {\bibinfo {author} {\bibfnamefont {W.~W.}\ \bibnamefont
  {Grabowski}}\ and\ \bibinfo {author} {\bibfnamefont {L.-P.}\ \bibnamefont
  {Wang}},\ }\href {\doibase 10.1146/annurev-fluid-011212-140750} {\bibfield
  {journal} {\bibinfo  {journal} {Annu. Rev. Fluid Mech.}\ }\textbf {\bibinfo
  {volume} {45}},\ \bibinfo {pages} {293} (\bibinfo {year} {2013})}\BibitemShut
  {NoStop}%
\bibitem [{\citenamefont {Pumir}\ and\ \citenamefont
  {Wilkinson}(2016)}]{Pumir2016}%
  \BibitemOpen
  \bibfield  {author} {\bibinfo {author} {\bibfnamefont {A.}~\bibnamefont
  {Pumir}}\ and\ \bibinfo {author} {\bibfnamefont {M.}~\bibnamefont
  {Wilkinson}},\ }\href {\doibase 10.1146/annurev-conmatphys-031115-011538}
  {\bibfield  {journal} {\bibinfo  {journal} {Annu. Rev. Condens. Matter
  Phys.}\ }\textbf {\bibinfo {volume} {7}},\ \bibinfo {pages} {141} (\bibinfo
  {year} {2016})}\BibitemShut {NoStop}%
\bibitem [{\citenamefont {Khain}\ and\ \citenamefont
  {Pinsky}(2018)}]{Khain2018}%
  \BibitemOpen
  \bibfield  {author} {\bibinfo {author} {\bibfnamefont {A.~P.}\ \bibnamefont
  {Khain}}\ and\ \bibinfo {author} {\bibfnamefont {M.}~\bibnamefont {Pinsky}},\
  }\href {\doibase 10.1017/9781139049481} {\emph {\bibinfo {title} {{Physical
  Processes in Clouds and Cloud Modeling}}}}\ (\bibinfo  {publisher} {Cambridge
  University Press},\ \bibinfo {year} {2018})\BibitemShut {NoStop}%
\bibitem [{\citenamefont {Zaichik}\ and\ \citenamefont
  {Alipchenkov}(2009)}]{Zaichik2009}%
  \BibitemOpen
  \bibfield  {author} {\bibinfo {author} {\bibfnamefont {L.~I.}\ \bibnamefont
  {Zaichik}}\ and\ \bibinfo {author} {\bibfnamefont {V.~M.}\ \bibnamefont
  {Alipchenkov}},\ }\href {\doibase 10.1088/1367-2630/11/10/103018} {\bibfield
  {journal} {\bibinfo  {journal} {New J. Phys.}\ }\textbf {\bibinfo {volume}
  {11}},\ \bibinfo {pages} {103018} (\bibinfo {year} {2009})}\BibitemShut
  {NoStop}%
\bibitem [{\citenamefont {Pan}\ and\ \citenamefont {Padoan}(2010)}]{pan2010}%
  \BibitemOpen
  \bibfield  {author} {\bibinfo {author} {\bibfnamefont {L.}~\bibnamefont
  {Pan}}\ and\ \bibinfo {author} {\bibfnamefont {P.}~\bibnamefont {Padoan}},\
  }\href {\doibase 10.1017/S0022112010002855} {\bibfield  {journal} {\bibinfo
  {journal} {J. Fluid Mech.}\ }\textbf {\bibinfo {volume} {661}},\ \bibinfo
  {pages} {73–107} (\bibinfo {year} {2010})}\BibitemShut {NoStop}%
\bibitem [{\citenamefont {Ireland}\ \emph {et~al.}(2016)\citenamefont
  {Ireland}, \citenamefont {Bragg},\ and\ \citenamefont
  {Collins}}]{ireland2016}%
  \BibitemOpen
  \bibfield  {author} {\bibinfo {author} {\bibfnamefont {P.~J.}\ \bibnamefont
  {Ireland}}, \bibinfo {author} {\bibfnamefont {A.~D.}\ \bibnamefont {Bragg}},
  \ and\ \bibinfo {author} {\bibfnamefont {L.~R.}\ \bibnamefont {Collins}},\
  }\href {\doibase 10.1017/jfm.2016.238} {\bibfield  {journal} {\bibinfo
  {journal} {J. Fluid Mech.}\ }\textbf {\bibinfo {volume} {796}},\ \bibinfo
  {pages} {617–658} (\bibinfo {year} {2016})}\BibitemShut {NoStop}%
\bibitem [{\citenamefont {Petersen}\ \emph {et~al.}(2019)\citenamefont
  {Petersen}, \citenamefont {Baker},\ and\ \citenamefont
  {Coletti}}]{petersen2019}%
  \BibitemOpen
  \bibfield  {author} {\bibinfo {author} {\bibfnamefont {A.~J.}\ \bibnamefont
  {Petersen}}, \bibinfo {author} {\bibfnamefont {L.}~\bibnamefont {Baker}}, \
  and\ \bibinfo {author} {\bibfnamefont {F.}~\bibnamefont {Coletti}},\ }\href
  {\doibase 10.1017/jfm.2019.31} {\bibfield  {journal} {\bibinfo  {journal} {J.
  Fluid Mech.}\ }\textbf {\bibinfo {volume} {864}},\ \bibinfo {pages}
  {925–970} (\bibinfo {year} {2019})}\BibitemShut {NoStop}%
\bibitem [{\citenamefont {Bertens}\ \emph {et~al.}(2021)\citenamefont
  {Bertens}, \citenamefont {Bagheri}, \citenamefont {Xu}, \citenamefont
  {Bodenschatz},\ and\ \citenamefont {Moláček}}]{Bertens2021}%
  \BibitemOpen
  \bibfield  {author} {\bibinfo {author} {\bibfnamefont {G.}~\bibnamefont
  {Bertens}}, \bibinfo {author} {\bibfnamefont {G.}~\bibnamefont {Bagheri}},
  \bibinfo {author} {\bibfnamefont {H.}~\bibnamefont {Xu}}, \bibinfo {author}
  {\bibfnamefont {E.}~\bibnamefont {Bodenschatz}}, \ and\ \bibinfo {author}
  {\bibfnamefont {J.}~\bibnamefont {Moláček}},\ }\href {\doibase
  10.1063/5.0065806} {\bibfield  {journal} {\bibinfo  {journal} {Rev. Sci.
  Instrum.}\ }\textbf {\bibinfo {volume} {92}},\ \bibinfo {pages} {125105}
  (\bibinfo {year} {2021})}\BibitemShut {NoStop}%
\bibitem [{\citenamefont {Falkovich}\ \emph {et~al.}(2002)\citenamefont
  {Falkovich}, \citenamefont {Fouxon},\ and\ \citenamefont
  {Stepanov}}]{Falkovich2002}%
  \BibitemOpen
  \bibfield  {author} {\bibinfo {author} {\bibfnamefont {G.}~\bibnamefont
  {Falkovich}}, \bibinfo {author} {\bibfnamefont {A.}~\bibnamefont {Fouxon}}, \
  and\ \bibinfo {author} {\bibfnamefont {M.~G.}\ \bibnamefont {Stepanov}},\
  }\href {\doibase 10.1038/nature00983} {\bibfield  {journal} {\bibinfo
  {journal} {Nature}\ }\textbf {\bibinfo {volume} {419}},\ \bibinfo {pages}
  {151} (\bibinfo {year} {2002})}\BibitemShut {NoStop}%
\bibitem [{\citenamefont {Wilkinson}\ and\ \citenamefont
  {Mehlig}(2005)}]{Wilkinson2005}%
  \BibitemOpen
  \bibfield  {author} {\bibinfo {author} {\bibfnamefont {M.}~\bibnamefont
  {Wilkinson}}\ and\ \bibinfo {author} {\bibfnamefont {B.}~\bibnamefont
  {Mehlig}},\ }\href {\doibase 10.1209/epl/i2004-10532-7} {\bibfield  {journal}
  {\bibinfo  {journal} {EPL}\ }\textbf {\bibinfo {volume} {71}},\ \bibinfo
  {pages} {186} (\bibinfo {year} {2005})}\BibitemShut {NoStop}%
\bibitem [{\citenamefont {Bec}\ \emph {et~al.}(2010)\citenamefont {Bec},
  \citenamefont {Biferale}, \citenamefont {Cencini}, \citenamefont {Lanotte},\
  and\ \citenamefont {Toschi}}]{bec2010}%
  \BibitemOpen
  \bibfield  {author} {\bibinfo {author} {\bibfnamefont {J.}~\bibnamefont
  {Bec}}, \bibinfo {author} {\bibfnamefont {L.}~\bibnamefont {Biferale}},
  \bibinfo {author} {\bibfnamefont {M.}~\bibnamefont {Cencini}}, \bibinfo
  {author} {\bibfnamefont {A.~S.}\ \bibnamefont {Lanotte}}, \ and\ \bibinfo
  {author} {\bibfnamefont {F.}~\bibnamefont {Toschi}},\ }\href {\doibase
  10.1017/S0022112010000029} {\bibfield  {journal} {\bibinfo  {journal} {J.
  Fluid Mech.}\ }\textbf {\bibinfo {volume} {646}},\ \bibinfo {pages}
  {527–536} (\bibinfo {year} {2010})}\BibitemShut {NoStop}%
\bibitem [{\citenamefont {Bewley}\ \emph {et~al.}(2013)\citenamefont {Bewley},
  \citenamefont {Saw},\ and\ \citenamefont {Bodenschatz}}]{Bewley2013}%
  \BibitemOpen
  \bibfield  {author} {\bibinfo {author} {\bibfnamefont {G.~P.}\ \bibnamefont
  {Bewley}}, \bibinfo {author} {\bibfnamefont {E.-W.}\ \bibnamefont {Saw}}, \
  and\ \bibinfo {author} {\bibfnamefont {E.}~\bibnamefont {Bodenschatz}},\
  }\href {\doibase 10.1088/1367-2630/15/8/083051} {\bibfield  {journal}
  {\bibinfo  {journal} {New J. Phys.}\ }\textbf {\bibinfo {volume} {15}},\
  \bibinfo {pages} {083051} (\bibinfo {year} {2013})}\BibitemShut {NoStop}%
\bibitem [{\citenamefont {Vo{\ss}kuhle}\ \emph {et~al.}(2014)\citenamefont
  {Vo{\ss}kuhle}, \citenamefont {Pumir}, \citenamefont {L{\'{e}}v{\^{e}}que},\
  and\ \citenamefont {Wilkinson}}]{Vosskuhle2014}%
  \BibitemOpen
  \bibfield  {author} {\bibinfo {author} {\bibfnamefont {M.}~\bibnamefont
  {Vo{\ss}kuhle}}, \bibinfo {author} {\bibfnamefont {A.}~\bibnamefont {Pumir}},
  \bibinfo {author} {\bibfnamefont {E.}~\bibnamefont {L{\'{e}}v{\^{e}}que}}, \
  and\ \bibinfo {author} {\bibfnamefont {M.}~\bibnamefont {Wilkinson}},\ }\href
  {\doibase 10.1017/jfm.2014.259} {\bibfield  {journal} {\bibinfo  {journal}
  {J. Fluid Mech.}\ }\textbf {\bibinfo {volume} {749}},\ \bibinfo {pages} {841}
  (\bibinfo {year} {2014})}\BibitemShut {NoStop}%
\bibitem [{\citenamefont {Bhatnagar}\ \emph {et~al.}(2022)\citenamefont
  {Bhatnagar}, \citenamefont {Pandey}, \citenamefont {Perlekar},\ and\
  \citenamefont {Mitra}}]{Bhatnagar2022}%
  \BibitemOpen
  \bibfield  {author} {\bibinfo {author} {\bibfnamefont {A.}~\bibnamefont
  {Bhatnagar}}, \bibinfo {author} {\bibfnamefont {V.}~\bibnamefont {Pandey}},
  \bibinfo {author} {\bibfnamefont {P.}~\bibnamefont {Perlekar}}, \ and\
  \bibinfo {author} {\bibfnamefont {D.}~\bibnamefont {Mitra}},\ }\href
  {\doibase 10.1098/rsta.2021.0086} {\bibfield  {journal} {\bibinfo  {journal}
  {Philos. Trans. Royal Soc. A}\ }\textbf {\bibinfo {volume} {380}},\ \bibinfo
  {pages} {20210086} (\bibinfo {year} {2022})}\BibitemShut {NoStop}%
\bibitem [{\citenamefont {Risius}\ \emph {et~al.}(2015)\citenamefont {Risius},
  \citenamefont {Xu}, \citenamefont {Di~Lorenzo}, \citenamefont {Xi},
  \citenamefont {Siebert}, \citenamefont {Shaw},\ and\ \citenamefont
  {Bodenschatz}}]{Risius2015}%
  \BibitemOpen
  \bibfield  {author} {\bibinfo {author} {\bibfnamefont {S.}~\bibnamefont
  {Risius}}, \bibinfo {author} {\bibfnamefont {H.}~\bibnamefont {Xu}}, \bibinfo
  {author} {\bibfnamefont {F.}~\bibnamefont {Di~Lorenzo}}, \bibinfo {author}
  {\bibfnamefont {H.}~\bibnamefont {Xi}}, \bibinfo {author} {\bibfnamefont
  {H.}~\bibnamefont {Siebert}}, \bibinfo {author} {\bibfnamefont {R.~A.}\
  \bibnamefont {Shaw}}, \ and\ \bibinfo {author} {\bibfnamefont
  {E.}~\bibnamefont {Bodenschatz}},\ }\href {\doibase 10.5194/amt-8-3209-2015}
  {\bibfield  {journal} {\bibinfo  {journal} {Atmos. Meas. Tech.}\ }\textbf
  {\bibinfo {volume} {8}},\ \bibinfo {pages} {3209} (\bibinfo {year}
  {2015})}\BibitemShut {NoStop}%
\bibitem [{\citenamefont {Meibohm}\ \emph {et~al.}(2021)\citenamefont
  {Meibohm}, \citenamefont {Pandey}, \citenamefont {Bhatnagar}, \citenamefont
  {Gustavsson}, \citenamefont {Mitra}, \citenamefont {Perlekar},\ and\
  \citenamefont {Mehlig}}]{Meibohm2021}%
  \BibitemOpen
  \bibfield  {author} {\bibinfo {author} {\bibfnamefont {J.}~\bibnamefont
  {Meibohm}}, \bibinfo {author} {\bibfnamefont {V.}~\bibnamefont {Pandey}},
  \bibinfo {author} {\bibfnamefont {A.}~\bibnamefont {Bhatnagar}}, \bibinfo
  {author} {\bibfnamefont {K.}~\bibnamefont {Gustavsson}}, \bibinfo {author}
  {\bibfnamefont {D.}~\bibnamefont {Mitra}}, \bibinfo {author} {\bibfnamefont
  {P.}~\bibnamefont {Perlekar}}, \ and\ \bibinfo {author} {\bibfnamefont
  {B.}~\bibnamefont {Mehlig}},\ }\href {\doibase
  10.1103/PhysRevFluids.6.L062302} {\bibfield  {journal} {\bibinfo  {journal}
  {Phys. Rev. Fluids}\ }\textbf {\bibinfo {volume} {6}},\ \bibinfo {pages}
  {L062302} (\bibinfo {year} {2021})}\BibitemShut {NoStop}%
\bibitem [{\citenamefont {Meibohm}\ \emph {et~al.}(2022)\citenamefont
  {Meibohm}, \citenamefont {Gustavsson},\ and\ \citenamefont
  {Mehlig}}]{Meibohm2022}%
  \BibitemOpen
  \bibfield  {author} {\bibinfo {author} {\bibfnamefont {J.}~\bibnamefont
  {Meibohm}}, \bibinfo {author} {\bibfnamefont {K.}~\bibnamefont {Gustavsson}},
  \ and\ \bibinfo {author} {\bibfnamefont {B.}~\bibnamefont {Mehlig}},\
  }\href@noop {} {\enquote {\bibinfo {title} {Caustics in turbulent aerosols
  form along vieillefosse line at weak particle inertia},}\ } (\bibinfo {year}
  {2022})\BibitemShut {NoStop}%
\bibitem [{\citenamefont {Fouxon}\ and\ \citenamefont
  {Lee}(2020)}]{Fouxon2020}%
  \BibitemOpen
  \bibfield  {author} {\bibinfo {author} {\bibfnamefont {I.}~\bibnamefont
  {Fouxon}}\ and\ \bibinfo {author} {\bibfnamefont {C.}~\bibnamefont {Lee}},\
  }\href {\doibase 10.1103/PhysRevE.101.061101} {\bibfield  {journal} {\bibinfo
   {journal} {Phys. Rev. E}\ }\textbf {\bibinfo {volume} {101}},\ \bibinfo
  {pages} {061101} (\bibinfo {year} {2020})}\BibitemShut {NoStop}%
\bibitem [{\citenamefont {Lifshits}\ \emph {et~al.}(1988)\citenamefont
  {Lifshits}, \citenamefont {Gredeskul},\ and\ \citenamefont
  {Pastur}}]{lifshits1988}%
  \BibitemOpen
  \bibfield  {author} {\bibinfo {author} {\bibfnamefont {I.~M.}\ \bibnamefont
  {Lifshits}}, \bibinfo {author} {\bibfnamefont {S.~A.}\ \bibnamefont
  {Gredeskul}}, \ and\ \bibinfo {author} {\bibfnamefont {L.~A.}\ \bibnamefont
  {Pastur}},\ }\href@noop {} {\emph {\bibinfo {title} {Introduction to the
  Theory of Disordered Systems}}}\ (\bibinfo  {publisher} {New York: Wiley},\
  \bibinfo {year} {1988})\BibitemShut {NoStop}%
\bibitem [{\citenamefont {Derevyanko}\ \emph {et~al.}(2007)\citenamefont
  {Derevyanko}, \citenamefont {Falkovich}, \citenamefont {Turitsyn},\ and\
  \citenamefont {Turitsyn}}]{Derevyanko2007}%
  \BibitemOpen
  \bibfield  {author} {\bibinfo {author} {\bibfnamefont {S.~A.}\ \bibnamefont
  {Derevyanko}}, \bibinfo {author} {\bibfnamefont {G.}~\bibnamefont
  {Falkovich}}, \bibinfo {author} {\bibfnamefont {K.}~\bibnamefont {Turitsyn}},
  \ and\ \bibinfo {author} {\bibfnamefont {S.}~\bibnamefont {Turitsyn}},\
  }\href {\doibase 10.1080/14685240701332475} {\bibfield  {journal} {\bibinfo
  {journal} {J. Turbul.}\ }\textbf {\bibinfo {volume} {8}},\ \bibinfo {pages}
  {N16} (\bibinfo {year} {2007})}\BibitemShut {NoStop}%
\bibitem [{\citenamefont {Saito}\ and\ \citenamefont
  {Gotoh}(2018)}]{Saito2018}%
  \BibitemOpen
  \bibfield  {author} {\bibinfo {author} {\bibfnamefont {I.}~\bibnamefont
  {Saito}}\ and\ \bibinfo {author} {\bibfnamefont {T.}~\bibnamefont {Gotoh}},\
  }\href {\doibase 10.1088/1367-2630/aaa229} {\bibfield  {journal} {\bibinfo
  {journal} {New J. Phys.}\ }\textbf {\bibinfo {volume} {20}},\ \bibinfo
  {pages} {023001} (\bibinfo {year} {2018})}\BibitemShut {NoStop}%
\bibitem [{\citenamefont {Li}\ \emph {et~al.}(2020)\citenamefont {Li},
  \citenamefont {Brandenburg}, \citenamefont {Svensson}, \citenamefont
  {Haugen}, \citenamefont {Mehlig},\ and\ \citenamefont
  {Rogachevskii}}]{Li2020}%
  \BibitemOpen
  \bibfield  {author} {\bibinfo {author} {\bibfnamefont {X.-Y.}\ \bibnamefont
  {Li}}, \bibinfo {author} {\bibfnamefont {A.}~\bibnamefont {Brandenburg}},
  \bibinfo {author} {\bibfnamefont {G.}~\bibnamefont {Svensson}}, \bibinfo
  {author} {\bibfnamefont {N.~E.~L.}\ \bibnamefont {Haugen}}, \bibinfo {author}
  {\bibfnamefont {B.}~\bibnamefont {Mehlig}}, \ and\ \bibinfo {author}
  {\bibfnamefont {I.}~\bibnamefont {Rogachevskii}},\ }\href {\doibase
  10.1175/JAS-D-19-0107.1} {\bibfield  {journal} {\bibinfo  {journal} {J.
  Atmos. Sci.}\ }\textbf {\bibinfo {volume} {77}},\ \bibinfo {pages} {337 }
  (\bibinfo {year} {2020})}\BibitemShut {NoStop}%
\bibitem [{\citenamefont {Saffman}\ and\ \citenamefont
  {Turner}(1956)}]{Saffman1956}%
  \BibitemOpen
  \bibfield  {author} {\bibinfo {author} {\bibfnamefont {P.~G.}\ \bibnamefont
  {Saffman}}\ and\ \bibinfo {author} {\bibfnamefont {J.~S.}\ \bibnamefont
  {Turner}},\ }\href {\doibase 10.1017/S0022112056000020} {\bibfield  {journal}
  {\bibinfo  {journal} {J. Fluid Mech.}\ }\textbf {\bibinfo {volume} {1}},\
  \bibinfo {pages} {16–30} (\bibinfo {year} {1956})}\BibitemShut {NoStop}%
\bibitem [{\citenamefont {Maxey}(1987)}]{Maxey1987}%
  \BibitemOpen
  \bibfield  {author} {\bibinfo {author} {\bibfnamefont {M.~R.}\ \bibnamefont
  {Maxey}},\ }\href {\doibase 10.1017/S0022112087000193} {\bibfield  {journal}
  {\bibinfo  {journal} {J. Fluid Mech.}\ }\textbf {\bibinfo {volume} {174}},\
  \bibinfo {pages} {441–465} (\bibinfo {year} {1987})}\BibitemShut {NoStop}%
\bibitem [{\citenamefont {Balkovsky}\ \emph {et~al.}(2001)\citenamefont
  {Balkovsky}, \citenamefont {Falkovich},\ and\ \citenamefont
  {Fouxon}}]{Balkovsky2001}%
  \BibitemOpen
  \bibfield  {author} {\bibinfo {author} {\bibfnamefont {E.}~\bibnamefont
  {Balkovsky}}, \bibinfo {author} {\bibfnamefont {G.}~\bibnamefont
  {Falkovich}}, \ and\ \bibinfo {author} {\bibfnamefont {A.}~\bibnamefont
  {Fouxon}},\ }\href {\doibase 10.1103/PhysRevLett.86.2790} {\bibfield
  {journal} {\bibinfo  {journal} {Phys. Rev. Lett.}\ }\textbf {\bibinfo
  {volume} {86}},\ \bibinfo {pages} {2790} (\bibinfo {year}
  {2001})}\BibitemShut {NoStop}%
\bibitem [{\citenamefont {Falkovich}\ and\ \citenamefont
  {Pumir}(2004)}]{Falkovich2004}%
  \BibitemOpen
  \bibfield  {author} {\bibinfo {author} {\bibfnamefont {G.}~\bibnamefont
  {Falkovich}}\ and\ \bibinfo {author} {\bibfnamefont {A.}~\bibnamefont
  {Pumir}},\ }\href {\doibase 10.1063/1.1755722} {\bibfield  {journal}
  {\bibinfo  {journal} {Phys. Fluids}\ }\textbf {\bibinfo {volume} {16}},\
  \bibinfo {pages} {L47} (\bibinfo {year} {2004})}\BibitemShut {NoStop}%
\bibitem [{Note1()}]{Note1}%
  \BibitemOpen
  \bibinfo {note} {The equations apply for all droplets sizes of interest
  provided that $\tau $ is considered as effective time, see e.g. \cite
  {Fouxon2015}}\BibitemShut {NoStop}%
\bibitem [{\citenamefont {Maxey}\ and\ \citenamefont
  {Riley}(1983)}]{Maxey1983}%
  \BibitemOpen
  \bibfield  {author} {\bibinfo {author} {\bibfnamefont {M.~R.}\ \bibnamefont
  {Maxey}}\ and\ \bibinfo {author} {\bibfnamefont {J.~J.}\ \bibnamefont
  {Riley}},\ }\href {\doibase 10.1063/1.864230} {\bibfield  {journal} {\bibinfo
   {journal} {Phys. Fluids}\ }\textbf {\bibinfo {volume} {26}},\ \bibinfo
  {pages} {883} (\bibinfo {year} {1983})}\BibitemShut {NoStop}%
\bibitem [{\citenamefont {Lalescu}\ \emph {et~al.}(2022)\citenamefont
  {Lalescu}, \citenamefont {Bramas}, \citenamefont {Rampp},\ and\ \citenamefont
  {Wilczek}}]{Lalescu2022}%
  \BibitemOpen
  \bibfield  {author} {\bibinfo {author} {\bibfnamefont {C.~C.}\ \bibnamefont
  {Lalescu}}, \bibinfo {author} {\bibfnamefont {B.}~\bibnamefont {Bramas}},
  \bibinfo {author} {\bibfnamefont {M.}~\bibnamefont {Rampp}}, \ and\ \bibinfo
  {author} {\bibfnamefont {M.}~\bibnamefont {Wilczek}},\ }\href@noop {}
  {\bibfield  {journal} {\bibinfo  {journal} {Comput. Phys. Commun.}\ }\textbf
  {\bibinfo {volume} {278}},\ \bibinfo {pages} {108406} (\bibinfo {year}
  {2022})}\BibitemShut {NoStop}%
\bibitem [{\citenamefont {Landau}\ and\ \citenamefont
  {Lifshitz}(2013)}]{Landau2013}%
  \BibitemOpen
  \bibfield  {author} {\bibinfo {author} {\bibfnamefont {L.~D.}\ \bibnamefont
  {Landau}}\ and\ \bibinfo {author} {\bibfnamefont {E.~M.}\ \bibnamefont
  {Lifshitz}},\ }\href@noop {} {\emph {\bibinfo {title} {{Quantum Mechanics:
  non-relastivistic theory}}}},\ \bibinfo {edition} {3rd}\ ed.\ (\bibinfo
  {publisher} {Elsevier},\ \bibinfo {year} {2013})\BibitemShut {NoStop}%
\bibitem [{\citenamefont {Weber}(1869)}]{Weber1869}%
  \BibitemOpen
  \bibfield  {author} {\bibinfo {author} {\bibfnamefont {H.}~\bibnamefont
  {Weber}},\ }\href@noop {} {\bibfield  {journal} {\bibinfo  {journal} {Math.
  Ann.}\ }\textbf {\bibinfo {volume} {1}},\ \bibinfo {pages} {1} (\bibinfo
  {year} {1869})}\BibitemShut {NoStop}%
\bibitem [{\citenamefont {Frisch}(1995)}]{Frisch1995}%
  \BibitemOpen
  \bibfield  {author} {\bibinfo {author} {\bibfnamefont {U.}~\bibnamefont
  {Frisch}},\ }\href {\doibase 10.1017/CBO9781139170666} {\emph {\bibinfo
  {title} {{Turbulence}}}}\ (\bibinfo  {publisher} {Cambridge University
  Press},\ \bibinfo {year} {1995})\BibitemShut {NoStop}%
\bibitem [{\citenamefont {Fouxon}\ \emph {et~al.}(2015)\citenamefont {Fouxon},
  \citenamefont {Park}, \citenamefont {Harduf},\ and\ \citenamefont
  {Lee}}]{Fouxon2015}%
  \BibitemOpen
  \bibfield  {author} {\bibinfo {author} {\bibfnamefont {I.}~\bibnamefont
  {Fouxon}}, \bibinfo {author} {\bibfnamefont {Y.}~\bibnamefont {Park}},
  \bibinfo {author} {\bibfnamefont {R.}~\bibnamefont {Harduf}}, \ and\ \bibinfo
  {author} {\bibfnamefont {C.}~\bibnamefont {Lee}},\ }\href {\doibase
  10.1103/PhysRevE.92.033001} {\bibfield  {journal} {\bibinfo  {journal} {Phys.
  Rev. E}\ }\textbf {\bibinfo {volume} {92}},\ \bibinfo {pages} {033001}
  (\bibinfo {year} {2015})}\BibitemShut {NoStop}%
\bibitem [{\citenamefont {Abramowitz}\ and\ \citenamefont
  {Stegun}(1964)}]{Abramowitz1964}%
  \BibitemOpen
  \bibfield  {author} {\bibinfo {author} {\bibfnamefont {M.}~\bibnamefont
  {Abramowitz}}\ and\ \bibinfo {author} {\bibfnamefont {I.~A.}\ \bibnamefont
  {Stegun}},\ }\href@noop {} {\emph {\bibinfo {title} {{Handbook of
  Mathematical Functions with Formulas, Graphs, and Mathematical Tables}}}}\
  (\bibinfo  {publisher} {U.S. Government Printing Office},\ \bibinfo {year}
  {1964})\BibitemShut {NoStop}%
\bibitem [{\citenamefont {Falkovich}\ \emph {et~al.}(2001)\citenamefont
  {Falkovich}, \citenamefont {Gaw\ifmmode~\mbox{\c{e}}\else \c{e}\fi{}dzki},\
  and\ \citenamefont {Vergassola}}]{Falkovich2001}%
  \BibitemOpen
  \bibfield  {author} {\bibinfo {author} {\bibfnamefont {G.}~\bibnamefont
  {Falkovich}}, \bibinfo {author} {\bibfnamefont {K.}~\bibnamefont
  {Gaw\ifmmode~\mbox{\c{e}}\else \c{e}\fi{}dzki}}, \ and\ \bibinfo {author}
  {\bibfnamefont {M.}~\bibnamefont {Vergassola}},\ }\href {\doibase
  10.1103/RevModPhys.73.913} {\bibfield  {journal} {\bibinfo  {journal} {Rev.
  Mod. Phys.}\ }\textbf {\bibinfo {volume} {73}},\ \bibinfo {pages} {913}
  (\bibinfo {year} {2001})}\BibitemShut {NoStop}%
\bibitem [{\citenamefont {Fouxon}(2012)}]{Fouxon2012}%
  \BibitemOpen
  \bibfield  {author} {\bibinfo {author} {\bibfnamefont {I.}~\bibnamefont
  {Fouxon}},\ }\href {\doibase 10.1103/PhysRevLett.108.134502} {\bibfield
  {journal} {\bibinfo  {journal} {Phys. Rev. Lett.}\ }\textbf {\bibinfo
  {volume} {108}},\ \bibinfo {pages} {134502} (\bibinfo {year}
  {2012})}\BibitemShut {NoStop}%
\bibitem [{\citenamefont {Fouxon}\ \emph {et~al.}(2022)\citenamefont {Fouxon},
  \citenamefont {Lee},\ and\ \citenamefont {Lee}}]{Fouxon2022}%
  \BibitemOpen
  \bibfield  {author} {\bibinfo {author} {\bibfnamefont {I.}~\bibnamefont
  {Fouxon}}, \bibinfo {author} {\bibfnamefont {S.}~\bibnamefont {Lee}}, \ and\
  \bibinfo {author} {\bibfnamefont {C.}~\bibnamefont {Lee}},\ }\href {\doibase
  10.48550/ARXIV.2205.06972} {\enquote {\bibinfo {title} {Intermittency and
  collisions of fast sedimenting droplets in turbulence},}\ } (\bibinfo {year}
  {2022})\BibitemShut {NoStop}%
\bibitem [{\citenamefont {Ishihara}\ \emph {et~al.}(2007)\citenamefont
  {Ishihara}, \citenamefont {Kaneda}, \citenamefont {Yokokawa}, \citenamefont
  {Itakura},\ and\ \citenamefont {Uno}}]{Ishihara2007}%
  \BibitemOpen
  \bibfield  {author} {\bibinfo {author} {\bibfnamefont {T.}~\bibnamefont
  {Ishihara}}, \bibinfo {author} {\bibfnamefont {Y.}~\bibnamefont {Kaneda}},
  \bibinfo {author} {\bibfnamefont {M.}~\bibnamefont {Yokokawa}}, \bibinfo
  {author} {\bibfnamefont {K.}~\bibnamefont {Itakura}}, \ and\ \bibinfo
  {author} {\bibfnamefont {A.}~\bibnamefont {Uno}},\ }\href {\doibase
  10.1017/S0022112007008531} {\bibfield  {journal} {\bibinfo  {journal} {J.
  Fluid Mech.}\ }\textbf {\bibinfo {volume} {592}},\ \bibinfo {pages}
  {335–366} (\bibinfo {year} {2007})}\BibitemShut {NoStop}%
\end{thebibliography}%

\onecolumngrid
\newpage

\setcounter{page}{1}
{
\centering

    {\large \textbf{Supplementary Information for \\``Quantitative prediction of sling events in turbulence at high Reynolds numbers''}}

    \vspace{.5cm}
    Tobias B\"atge, Itzhak Fouxon and Michael Wilczek\\
    (Dated: \today)
    \vspace{.5cm}
    
}

\renewcommand{\figurename}{Supplementary Figure}
\renewcommand{\tablename}{Supplementary Table}
\renewcommand{\thesection}{\arabic{section}}
\setcounter{table}{0}
\renewcommand{\thetable}{S\arabic{table}}%
\setcounter{figure}{0}
\renewcommand{\thefigure}{S\arabic{figure}}%
\setcounter{equation}{0}
\def\theequation{S\arabic{equation}}
\setcounter{section}{0}

The so-called sling effect corresponds to caustics in the flow of inertial particles. The occurrence of caustics is equivalent to a singularity in the particle velocity gradient. In the main text, we have shown that this can be reduced to a one-dimensional problem considering blow-ups of the minimal real eigenvalue, $\sigma$ of the particle velocity gradient. As we will detail below, those coincide with zeros of the "wavefunction" $\psi$ that obeys a one-dimensional Schr\"odinger equation with a certain value of energy, see Eq.~($5)$ in the main text.
In section \ref{sec:globalsolutionpsi}, we  discuss the general solution for $\psi$. In the main text, we use this solution to develop a criterion for the occurrence of sling events along particle trajectories. In section \ref{sec:deterministicdensityofstates}, we describe the results from the perspective of $1d$ localization problem along the lines of \cite{lifshits1988}. In section \ref{sec:parameterstimation}, we provide more details on the parameters for our prediction of the sling rate. Finally, in section \ref{sec:S4}, we discuss the implications to collision rates in clouds.

\section{Global solution for $\psi$}
\label{sec:globalsolutionpsi}
In the main text, we show that sling events can be described by the dynamics of the most negative eigenvalues of velocity gradients in form of blowups. With a change of variables one arrives at Schr\"odinger equation (Eq.~(5) in the main text).
Here, we demonstrate how the general solution of the Schr\"odinger equation 
\begin{equation}
-\frac{d^2\psi}{dt^2}+\frac{a}{\tau}\psi=-\frac{\psi}{4\tau^2}, \label{eq:Sa}    
\end{equation}
can be constructed. The equation is characterized by the turning points where the "potential" $a/\tau$ equals the "energy" $-1/(4\tau^2)$. These points, where $a=-1/(4\tau)$, are rare by the assumption of small Stokes number. Thus most of the time the particle is found in classically forbidden regions where the potential is larger than the energy, $a>-1/(4\tau)$. In these regions the quasi-classical solution (Eq.$\ (6)$ in the main text) 
\begin{equation}
\psi(t) \sim \frac{\exp\left(\int^t p(s) ds\right)}{\sqrt{p(t)}},\ \
p(t)\equiv \sqrt{\frac{1}{4\tau^2}\!+\!\frac{a(t)}{\tau}}, \label{eq:Squ}    
\end{equation}
applies, providing for exponential behavior of $\psi$. In contrast, in the rare, classically allowed regions with $a>-1/(4\tau)$ the wavefunction oscillates. This is where the zeros of the wavefunction that correspond to the sling events occur. These regions occur due to rare excursions of $a(t)$ to values $-1/(4\tau)-\Delta\tau$ with $\Delta>0$. Since the excursions are rare, it is most probable that after an excursion, the random process $a(t)$ returns to the range $|a|\ll 1/\tau$. The resulting time series of $a(t)$ is characterized by a sharp minimum around which we can use the quadratic expansion, see Fig.~1 in the main text,
\begin{align} 
\frac{a}{\tau}+\frac{1}{4\tau^2}\approx-\Delta+\frac{\Delta}{D^2}t^2.
\label{eq:Sas}
\end{align}
Using the above form of the potential in Eq.~(\ref{eq:Sa}) results in 
the Weber equation (Eq.$\ (8)$ in the main text) whose solutions are well-known. These are given by a combination of parabolic cylinder functions $U$ and $V$
\begin{align} 
\psi=c_1 U\left(-\frac{\Delta}{\kappa^2}, -\kappa t\right)+c_2V\left(-\frac{\Delta}{\kappa^2}, -\kappa t\right);\ \ \ \ 
\kappa^4\equiv \frac{4\Delta}{D^2}, \label{Sgenea}
\end{align}
where $c_i$ are constants, see e.g. Chapter $19$ in \cite{Abramowitz1964}. We defined above $\kappa^2$ via $\Delta$ and $D$ introduced in the main text. Thus, we have forms of the solution both in the classically forbidden and allowed regions, and what remains is to match these forms in the intermediate regions. 

To describe the intermediate regions, we consider returns of  $a(t)$ after a rare excursion below $-1/(4\tau)$ to the range of $|a|\ll 1/\tau$. We assume that the return can be modeled by extending the parabolic approximation given by Eq.~(\ref{eq:Sas}) up to the domain with $\kappa^4t^2\gg \Delta$. In that domain, both Eqs.~(\ref{eq:Squ}) and (\ref{Sgenea}) hold which allows to perform the matching and fix $c_1$ and $c_2$ (we use that rare excursions are well-isolated from each other). We have in the range of $t$ larger than some $-T$ with $T>0$ from Eq.~(\ref{eq:Squ}) that
\begin{align}
\psi(t)\approx \psi(-T)\sqrt{\frac{p(-T)}{p(t)}}\exp\left(\int_{-T}^t dt'p(t')\right).
\end{align}
This becomes in the range $\kappa^4t^2\gg \Delta$
\begin{align} 
\psi\approx \psi(-T)\sqrt{\frac{T}{|t|}}\exp\left(\int_{-T}^t dt'\sqrt{-\Delta+\frac{\kappa^4t^2}{4}}\right)=&\psi(-T)\sqrt{\frac{T}{|t|}}\exp\left(\frac{\kappa^2}{2}\int_{-T}^t dt'\left(|t|-\frac{2\Delta}{\kappa^4|t|}\right)\right)\nonumber\\
=&\psi(-T)\exp\left(\frac{\kappa^2(T^2-t^2)}{4}\right)\left(\frac{|t|}{T}\right)^{\Delta \kappa^{-2}-1/2}.
\end{align}
We compare this with the asymptotic expansions at large argument $V\propto \exp(\kappa^2 t^2/4)$ and \cite{Abramowitz1964} 
\begin{align} 
U\left(-\frac{\Delta}{\kappa^{2}}, -\kappa t\right)\approx \exp\left(-\frac{\kappa^2 t^2}{4}\right)|\kappa t|^{\Delta\kappa^{-2}-1/2}.
\end{align}
Demanding matching of Eq.~(\ref{Sgenea}), we find that in the region where the parabolic expansion of $a(t)$ near the minimum applies we have 
\begin{align} 
\psi=\psi(-T)\exp\left(\frac{\kappa^2 T^2}{4}\right)\left(\kappa T\right)^{1/2-\Delta\kappa^{-2}} U\left(-\frac{\Delta}{\kappa^2}, -\kappa t\right).
\end{align}
We conclude that the zeros of the solution coincide with zeros of $U$ above. The general properties of these zeros, described in Chapters $12$ and $19$ in \cite{Abramowitz1964}, imply the results described in the main text.
\section{Deterministic density of states in random potential}
\label{sec:deterministicdensityofstates}

\begin{figure}
    \centering
    \includegraphics{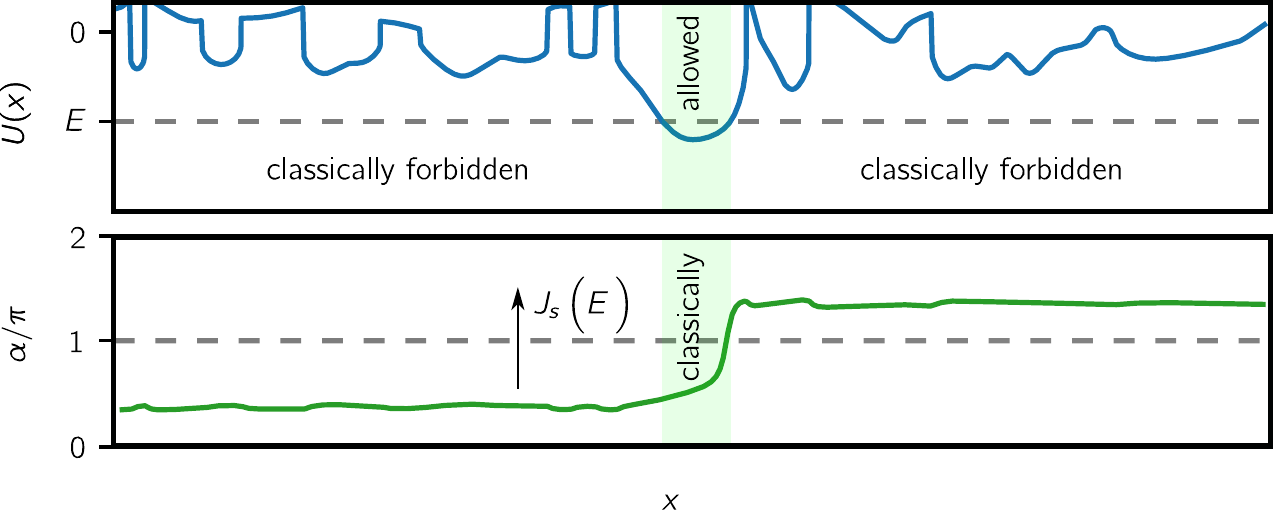}
    \caption{Analogy of sling dynamics (see also Fig.~2 in the main text) to 1d-localization in a random potential. By interpreting our time variable as space variable, $t \rightarrow x$, the velocity gradient $a$ as potential $U$, as well as the threshold value $-1/(4\tau)$ as the energy, we establish an analogy to a stationary 1d-Schr\"odinger equation. The probability flux $J_s$ between multiples of $\pi$ for the phase $\alpha$ corresponds to the sling rate. Note that the sudden jumps in $U$ are artifacts originating from points where the minimal real eigenvalue $a$ becomes complex. Those, however, do not lead to similar jumps in the dynamical variables such as $\alpha$. }
    \label{fig:S1}
\end{figure}
In the following, we will discuss the implications of our work to the $1d$ localization problem. We introduced $\psi$ where its zeros capture the occurrence of sling events. Coincidentally, $\psi$ is equivalent to the wavefunction of the one-dimensional Schr\"odinger equation. However, if a wavefunction has $n$ zeros on a certain interval, then it corresponds to the $n-$th energy level with suitable boundary conditions (here the ground state is traditionally considered as zeroth level). This establishes a connection to counting the energy levels, or equivalently the density of states, of the Schr\"odinger equation with a random potential, a problem well-studied in the context of one-dimensional localization problems \cite{lifshits1988}. From this viewpoint, we calculate the density of states at large negative energies in the main text (which apparently was not previously done in the literature covering the localization problem). \\
Here we consider the time-independent one-dimensional Schr\"odinger equation with random potential $U(x)$
\begin{align}
-\psi''+U\psi=E\psi,
\end{align}
where $\psi(x)$ is the wavefunction and $E$ is the energy. 
This is equivalent to Eq.~($5$) in the main text where time $t$ has effectively the role of space $x$ and $a(t)$ is corresponding to the potential $U(x)$. 
In Fig.~\ref{fig:S1}, we show the corresponding quantities. There it becomes transparent that the excursions in main text below the threshold are in this picture classically allowed regions i.e. where the energy is greater than the potential.

The Schr\"odinger equation is solved on the interval $(0, L)$. The boundary conditions are irrelevant in the sense that it can be demonstrated by using the variational principle that in the thermodynamic limit $L\to \infty$ all boundary conditions produce the same macroscopic properties such as the density of states. We will use the set of conditions provided by fixing values of the logarithmic derivatives of $\psi$ at both ends of the interval, $\psi'(0)/\psi(0)$ and $\psi'(L)/\psi(L)$, respectively.

Alternatively, we can consider the Schr\"odinger equation as a linear evolution of a two-dimensional vector $\bm r$ according to
\begin{align}
\bm r\equiv \begin{pmatrix}
\psi\\
\psi'
\end{pmatrix},\ \ \sigma\equiv \begin{pmatrix}
0 & 1\\
U(x)-E & 0
\end{pmatrix},\ \ \frac{d\bm r}{dx}=\sigma\bm r \, .\label{Sres}
\end{align}
If $x$ plays the role of time, this equation describes the distance $\bm r$ between two infinitesimally close trajectories moving in the flow with local matrix of velocity gradients $\sigma$.
This allows to use concepts such as Lyapunov exponents that were considered extensively in the context of turbulent transport, see e.g. \cite{Falkovich2001} and references therein. 
Here, similarly to the study of the Lyapunov exponent, it is useful to introduce polar coordinates in the $\bm r-$plane via $\psi'=r\cos\alpha$ and $\psi=r\sin\alpha$. Eq.~(\ref{Sres}) then becomes
\begin{align}
\frac{d\ln r}{dx}=(U-E+1)\sin\alpha \cos\alpha,\ \ \frac{d\alpha}{dx}=\cos^2\alpha+(E-U(x))\sin^2\alpha\equiv \Phi(E, \alpha(x), U(x)),\label{Sse}
\end{align}
where we introduced the function $\Phi$. We designate the value of the prescribed logarithmic derivative $\psi'(0)/\psi(0)$ by $\cot \alpha_0$ with $0\leq \alpha_0<\pi$. Then $\alpha(x)$ and the logarithmic derivative of $r$ are uniquely determined  by $\alpha_0$ via the above first-order ordinary differential equation. The phase considered as a function of $E$ at some fixed $\alpha_0$ obeys
\begin{align}
\frac{\partial}{\partial x} \frac{\partial \alpha(x, E)}{\partial E}=\sin^2\alpha+\left(E-U(x)-1\right)\sin 2\alpha \frac{\partial \alpha(x, E)}{\partial E},
\end{align}
obtained by differentiating the last of Eqs.~(\ref{Sse}). The solution of this equation gives
\begin{align}
\frac{\partial \alpha(x, E)}{\partial E}=\int_0^x \sin^2\alpha(y)\exp\left(\int_y^x \left(E-U(x')-1\right)\sin 2\alpha(x')dx'\right)dy>0.
\end{align}
We consider construction of eigenfunctions on the interval $(0, L)$ with $\alpha(0)=\alpha_0$ and $\alpha(x=L)=\alpha_L$ where $0\leq \alpha_L<\pi$. The solution of the last of Eqs.~(\ref{Sse}) at given $\alpha_0$ defines the value of $\alpha(x=L)$ as a function of $E$ that we write as $\alpha_L(E)$. The above equation shows that $\alpha_L(E)$ is a monotonically increasing function of $E$. The eigenvalues are determined by the condition $\alpha_L(E_n)=\alpha_L+\pi n$. Therefore, the number of states with energy between some $E_1$ and $E_2$ is the number of multiples of $\pi$ in $\alpha_L(E_2)-\alpha_L(E_1)$ (here $E_2>E_1$).
We conclude that density of states per unit length $\rho(E)$ obeys
\begin{align}
\int_{E_1}^{E_2}\rho(E)dE=\frac{1}{\pi L}\left[\alpha_L(E_2)-\alpha_L(E_1)\right]=\frac{1}{\pi L}\left\lfloor\int_0^L \Phi(E_2, \alpha(x), U(x))dx-\int_0^L \Phi(E_1, \alpha(x), U(x))dx\right\rfloor,\label{Sst}
\end{align}
where we used $\Phi(E, \alpha(x), U(x))$ defined in Eq.~(\ref{Sse}). We observe that the process $\Phi(E, \alpha(x), U(x))$ becomes stationary (in $x$) at large $x$ because the reduced
phase $\alpha_r$, which is defined by $0\leq \alpha_r<\pi$ and $\alpha(x)=\alpha_r+\pi k$ with integer $k$, does. We find from the above equation in the limit $L\to\infty$
\begin{align}
\int_{E_1}^{E_2}\rho(E)dE=\frac{\langle \Phi(E_2, \alpha, U)\rangle-\langle \Phi(E_1, \alpha, U)\rangle}{\pi},\label{Ssto}
\end{align}
where the averaging must be performed in the steady state. This formula shows that $\rho(E)$ is independent of the realization of the potential at $L\to\infty$. 

The average in Eq.~(\ref{Ssto}) can be studied by considering the PDF of the reduced phase
\begin{align}
P(\alpha_r, x)\!\equiv \!\sum_{k=-\infty}^{\infty} \left\langle \delta\left(\alpha(x)\!-\!\alpha_r\!-\!k\pi\right)\right\rangle,\ \ \frac{\partial P}{\partial x}\!+\!\frac{\partial J}{\partial \alpha_r}\!=\!0,\ \
J(\alpha_r)\!\equiv \!\sum_{k=-\infty}^{\infty} \left\langle \Phi(E, \alpha(x), U(x))\delta\left(\alpha(x)\!-\!\alpha_r\!-\!k\pi\right)\right\rangle,
\end{align}
where $J$ is the probability currrent. Since the PDF becomes ''stationary" at large $x$, i.e. $x$-independent, $J$ becomes independent of $\alpha_r$. We conclude that the stationary current $J_s$ obeys
\begin{align}
J_s=\frac{1}{\pi}\int_0^{\pi}J_s(\alpha_r)d\alpha_r= \frac{\left\langle \Phi(E, \alpha(x), U(x))\right\rangle}{\pi},
\end{align}
where the averaging must be done in the steady state. We find from Eq.~(\ref{Sst}) that
\begin{align}
\int_{E_1}^{E_2}\rho(E)dE=J_s(E_2)-J_s(E_1). \label{Sas}
\end{align}
In cases of interest here, $\rho(E)$ vanishes at large negative energies fast and the above equation gives
\begin{align}
\int_{-\infty}^{E}\rho(E')dE'=J_s(E). 
\end{align}
We observe that the left-hand side of the above equation is the number of zeros of the wavefunction per unit length at energy $E$. In Fig.~\ref{fig:S1}, we show how one potential minimum leading to a phase shift by $\pi$ i.e.~a zero of the wavefunction. This would be equivalent to a sling event in the main text. 
So from this viewpoint, the frequency of sling events $F$ corresponds one to one to the current $J_s(E=-1/(4\tau^2))$.
Since $\tau$ is only constrained by the condition that the Stokes number is small, we conclude that in the main text we derived the density of states for our random potential at large negative energies. 

\section{Estimating parameters for prediction}
\label{sec:parameterstimation}

\begin{figure}
    \centering
    \includegraphics[width=0.9\textwidth]{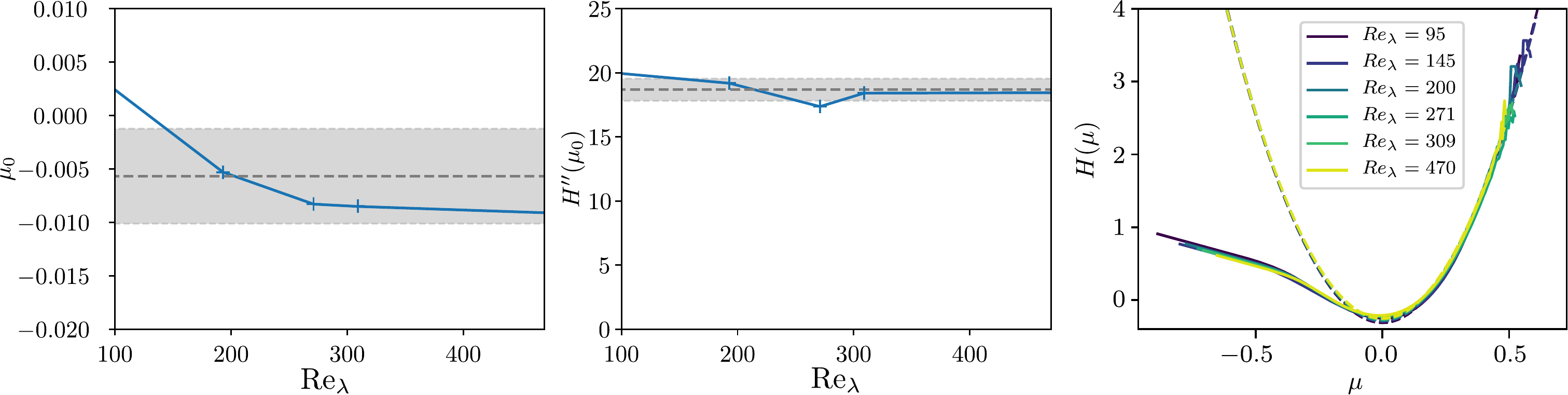}
    \caption{(Left, center) $\mu_0$ and $H''(\mu_0)$ depend on the Reynolds number only weakly. $\mu_0$ is obtained by taking the location of the minimum of the numerically evaluated large-deviation forms.  $H''(\mu_0)$ (center) is determined by computing the variance over the right tail. The gray dashed areas indicate the standard deviation. (Right) The numerically obtained large-deviation form and in dashed the corresponding log-normal approximation.}
    \label{fig:S2}
\end{figure}
In the main text, we split the sling rate $F$ into the frequency of excursions below a critical threshold $f$ times the probability that those excursions fulfill the criterion for a sling event to occur, $P(D\sqrt{\Delta}>1)$:
\begin{align}
    F=fP\left(D\sqrt{\Delta}>1\right)\,.
\end{align}
Our main interest is in getting insight into the above quantities that would allow to predict the values at the high Reynolds numbers characteristic of clouds. This requires the fitting of several parameters which we will explain in more detail in the following. \

\begin{figure}
    \centering
    \includegraphics[width=\textwidth]{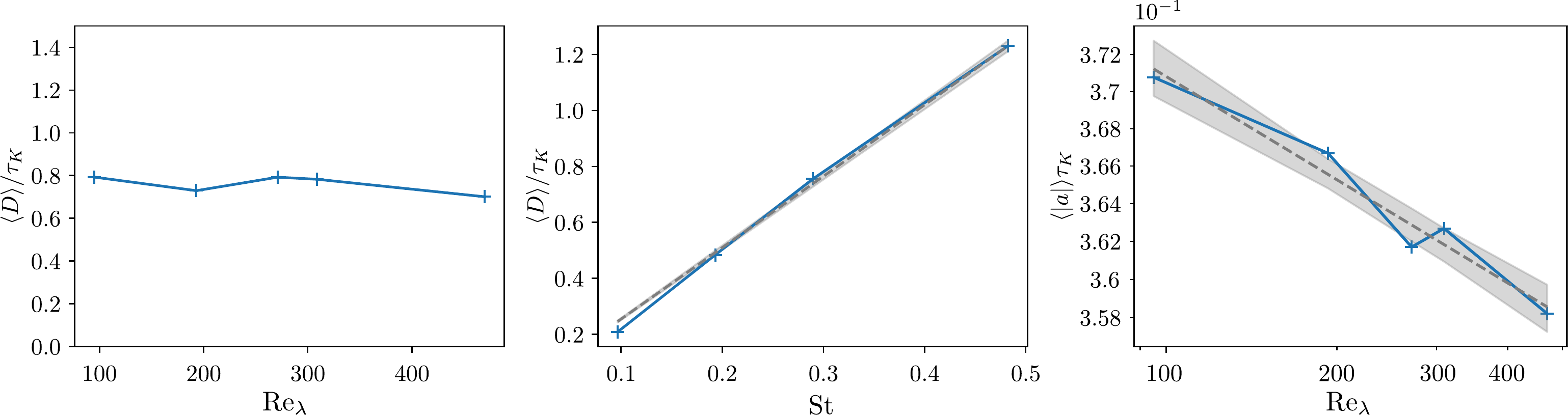}
    \caption{(Left) The mean duration of the excursions for different Reynolds numbers and St=0.3 shows only a weak Reynolds number dependence. (Center) The mean duration for fixed Reynolds number ($\text{Re}_\lambda=271$) increases approximately linearly with the Stokes number. The gray dashed line corresponds to linear fit $\langle D\rangle=b \text{St}\tau_K$ with $b=2.55\pm0.04$. (Right) The mean of the most negative real eigenvalue of the velocity gradient tensor of the flow $|a|$ (blue line) for different Reynolds numbers can be approximated by a power law. The gray dashed line corresponds to the fit used for our prediction and the gray shaded area shows its range of uncertainty. This corresponds to the confidence intervals of 75\% obtained by a bootstrap.}
    \label{fig:S3}
\end{figure}

We will begin with the excursion frequency. We decompose it as the fraction of time spent below the threshold divided by the mean duration of the excursion:
\begin{align}
f= \frac{1}{2\langle D\rangle}\int_{-\infty}^{-(4\tau)^{-1}}P(a)da\,.\label{eq:Sf}
\end{align}
In the main text, we argue that the mean duration is only weakly dependent on the Reynolds number but has a strong Stokes number dependence and should be approximately proportional to it, $\langle D\rangle=b\text{St}\tau_K$. Fig.~\ref{fig:S3} shows that our numerical data supports those claims, obtaining a value of $b=2.55\pm0.04$ through a fit.

To evaluate the integral in Eq.~(\ref{eq:Sf}), the large-deviation form and a log-normal approximation can be used for $P(a)$. In recent works \cite{Fouxon2020}, it has been shown that the empirically observed power-law behavior with Reynolds number (for $\text{Re}_\lambda>100$) of the dissipation rate moments implies the existence of a large-deviation function:
\begin{align}
    \mu_\epsilon=\frac{\ln (\epsilon/\langle\epsilon\rangle)}{\ln\text{Re}_\lambda}\qquad\text{with}\qquad    \lim_{\text{Re}_\lambda\rightarrow\infty}P(\mu_\epsilon)\propto\text{Re}_\lambda^{H_\epsilon(\mu_\epsilon)},
\end{align} where $H_\epsilon(\mu_\epsilon)$ is independent of Reynolds number. The formal condition of validity of the large-deviation theory is stringent, $\ln{\text{Re}_\lambda}\gg 1$. In practice, the theory often holds already for moderate Reynolds number, $\text{Re}_\lambda\gtrsim 100$. Once the shape of $H_\epsilon(\mu_\epsilon)$ is known, this can be used to express the PDF of $\epsilon$ for arbitary Reynolds number. 

The dissipation rate is a second-order quantity of the velocity gradients while $a$ is a first-order quantity. We heuristically argue that an extreme event in $a$ should also be dominating the dissipation rate. In this scenario $a$ would be in close relation to the $1/2$-moment of the dissipation rate. This motivates to use the same large-deviation approach also to describe $P(a)$, where in this context $\mu$ is defined as:
\begin{align}
    \mu=\frac{\ln (|a|/\langle |a|\rangle)}{\ln\text{Re}_\lambda}\qquad\text{with}\qquad    \lim_{\text{Re}_\lambda\rightarrow\infty}P(\mu)\propto\text{Re}_\lambda^{H(\mu)}.
\end{align}
Indeed, the simulation results show then a consistent large-deviation form across the studied Reynolds number range, see Fig.~\ref{fig:S2}.
To compute $H(\mu)$ from numerical data, we first numerically sample $P(\mu)$ and then invert the relation $P(\mu)\propto \text{Re}_\lambda^{-H(\mu)}$:
\begin{align}
    H(\mu)=\frac{\ln P(\mu)}{\ln\text{Re}_\lambda}+\mathrm{const}\,.
\end{align}
Here, the constant is given by normalization of $P(\mu)$.
 Once characterized, $H(\mu)$ defines the Reynolds dependence of $P(a)$ and in particular also the integral of interest in Eq.~\eqref{eq:Sf}.
To this end, the integration range covers only strongly negative $a$. This corresponds to the right tail of $H(\mu)$ only which can be in good approximation described by a parabola. This parabola is described by its position of the minimum $\mu_0$ and its second derivative at the minimum, $H''(\mu_0)$. 
To determine the minimum $\mu_0$, we can look at its position for different Reynolds numbers. 
In Fig.~\ref{fig:S2}, we show the inferred values for $\mu_0$, $H''(\mu_0)$ and the corresponding resulting log-normal approximation compared to numerical data. 
 Taking the mean as the value for $\mu_0$ and the standard deviation as error, we obtain $\mu_0\approx-0.006\pm0.005$. For all covered Reynolds numbers, $\mu_0$ is a small value is very close to zero.  
 Knowing where the minimum is, we can also calculate the variance relative to $\mu_0$ only using the right tail: $\langle (\mu-\mu_0)^2|\mu>\mu_0\rangle$. In accordance with the log-normal approximation, we find the expression for the corresponding $H''(\mu_0)=18.6\pm0.8$:
\begin{align}
     H''(\mu_0)=\lim_{\text{Re}_\lambda\to\infty}
    \frac{1}{\ln \text{Re}_\lambda\langle (\mu-\mu_0)^2|\mu>\mu_0\rangle}
\end{align}
To translate this back in terms of $a$ we need to characterise the mean of $|a|$ as it enters the definition of $\mu$. 
For the mean of $|a|$, we expect that in Kolmogorov units it decreases with a weak power-law dependence on Reynolds similar to the 1/2-moment of dissipation rate. In fact one can reasonably fit a power-law $\langle |a|\rangle=a_0\text{Re}_\lambda^{-\beta}$ to our numerical data as shown in Fig.~\ref{fig:S3}. The fit suggests $a_0=0.41\pm0.01$ and $\beta=0.022\pm0.003$.

With all of this we can express the tail of strongly negative $a$ of $P(a)$. Due to the log-normal approximation the integral can be captured by a complementary error function:
\begin{align}
\int_{-\infty}^{-\frac{1}{4\tau}}P(a)da=\frac{1}{2}\mathrm{erfc}\left(\sqrt{\frac{H''(\mu_0)}{2\ln\text{Re}_\lambda}}\left[\ln\left(\frac{\text{Re}_\lambda^{\beta}}{4a_0\text{St}}\right)-\mu_0\ln \text{Re}_\lambda\right]\right)\,.\label{eq:Sparameterization2}
\end{align}
This formula for the integral provides the left tail of the cumulative probability distribution 
function of $a$. Four empirical constants, that are fixed from the simulations, fix the full dependence on
the Reynolds and Stokes numbers.
Combined with the expression for the mean duration of excursions, this gives a prediction about the excursion frequency as function of $\text{St}$ and $\text{Re}_\lambda$ via five empirical constants.

\begin{figure}
    \centering
    \includegraphics[width=0.45\textwidth]{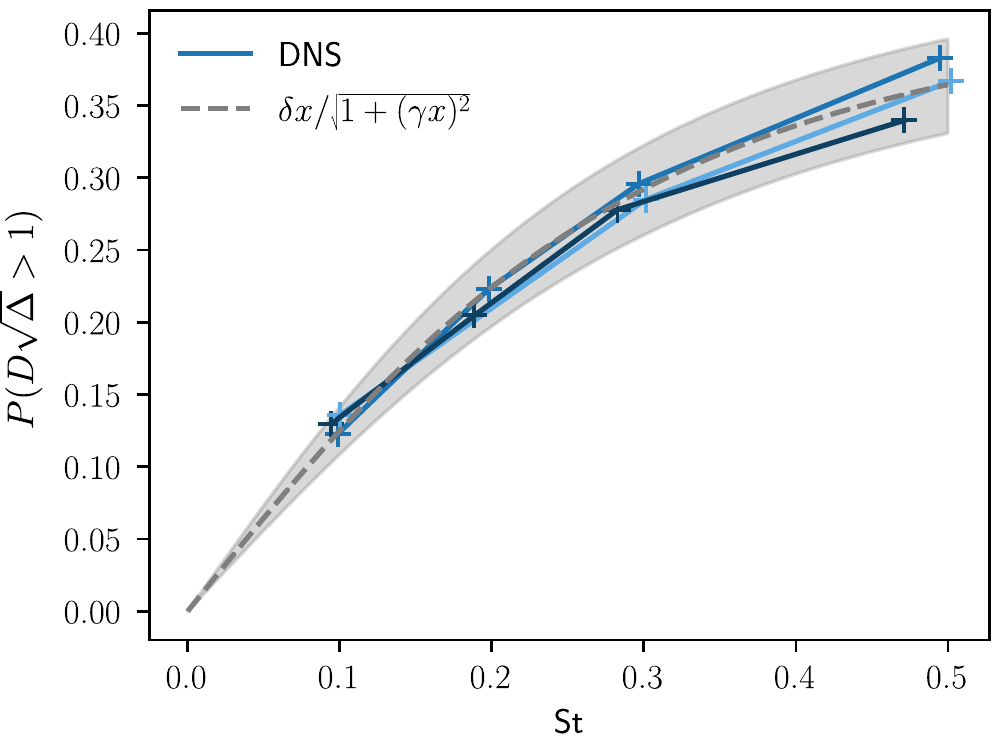}
    \caption{A sigmoid function can describe the Stokes number dependence of the probability that an excursion fulfills the criterion. The three solid line correspond to the Reynolds numbers $\text{Re}_\lambda = 193, 271, 470$ going from light to dark blue, respectively. The dashed gray line denotes the sigmoid fit where the gray area show the uncertainty range, obtained through a bootstrap corresponding to confidence intervals.}
    \label{fig:S4}
\end{figure}

Let us turn to the probability that given that an excursion is below $-1/(4\tau)$, it would actually lead to a sling event. In the main text, we argue that it should be roughly independent of Reynolds number. Also, we propose to capture it by an odd sigmoid function since this is the simplest way to obey the conditions of monotonous increase, vanishing at zero and finite limit at $St\to\infty$. Here we choose:
\begin{align}
    P(D\sqrt{\Delta}>1) = \delta\frac{ \text{St}}{\sqrt{1+\gamma \text{St}^2}}\,.
\end{align}
In Fig.~\ref{fig:S4}, we show the Stokes number dependence for different Reynolds numbers. Firstly, one cannot identify a clear Reynolds number trend. To determine the Stokes number dependence, we fit the above sigmoid function. This fit suggests $\delta = 1.30\pm0.05$ and $\gamma=8.8\pm1.2$.
Note that also other sigmoid functions can be used while the overall prediction of sling frequency remains consistent. The used function gives $0.44$ at $\text{St}=\infty$. This is the probability that a fluctuation of $a$ after becoming negative, will stay negative during time $2D$ and will have the minimum $-\Delta$ obeying $D\sqrt{\Delta}>1$. It is beyond our purposes here to check this prediction directly, however it might be relevant for future work.

Based on these fits, we obtain a prediction for the sling frequency in terms of the Reynolds and Stokes numbers using Eq.~(9) in the main text:
\begin{align}
F\tau_K=\frac{ \delta}{4b\,\sqrt{1+\gamma \text{St}^2}}\mathrm{erfc}\left(\sqrt{\frac{H''(\mu_0)}{2\ln\text{Re}_\lambda}}\left[\ln\left(\frac{\text{Re}_\lambda^{\beta}}{4a_0\text{St}}\right)-\mu_0\ln \text{Re}_\lambda\right]\right)\,.
\end{align}
This is the formula used in Fig.~4 in the main text to extrapolate our prediction to a wide range of Stokes and Reynolds numbers. 

We wish to consider $F$ at $\text{Re}_\lambda=10^4$ characteristic of clouds. Using the numerical values of the constants, we obtain  
\begin{align}&&
F=\frac{ 14.77}{\sqrt{1+8.8 \text{St}^2}}\mathrm{erfc}\left(-1.01\ln \text{St}-0.24\right). \label{Sfre}
\end{align}

\section{implications for clouds}
\label{sec:S4}
We study implications of our findings for the volume-averaged rate of droplet collisions $\Gamma$ at Reynolds numbers present in clouds. We consider collisions of equal-size droplets with radius $R$ and $\text{St}\ll 1$. We study the geometric collision kernel, where hydrodynamic interactions of the colliding particles are disregarded. The interactions could be then included via a multiplicative collision efficiency factor, see e.g. \cite{Falkovich2002}. In the following, we make a rough estimate of the relevance of slings to collisions based on our predicted sling rate. Along the way, we expose open research questions regarding the spatio-temporal structure of sling events as well as particle clustering properties.

Independently of $\text{Re}_{\lambda}$, after transients of order $\tau$, space can be decomposed into a region where particle velocities are unique differentiable functions of the particles' positions $\bm x(t)$, and the region of slings where the velocity is multi- (typically three- or five-) valued. The volume fraction of the former region is close to $1$ due to $\text{St}\ll 1$. In this region, the particle velocity $\bm v$ obeys, by definition, $\bm v=\bm v(\bm x(t), t)$, where $\bm v(\bm x, t)$ is a differentiable field. The effective flow of the particles $\bm v(\bm x, t)$ is compressible and in the region of the flow the particles collide as tracers in $\bm v(\bm x, t)$. We designate the corresponding collision rate by $\Gamma_{\text{tr}}$. In the sling region, vortices have time scales smaller of comparable with $\tau$ (more precisely these vortices either exist there currently, or existed in recent, not more than order $\tau$, prehistory of some of particles in the region). We designate the rate of collisions in the sling region by $\Gamma_{\text{sl}}$, so that the total rate of collisions in the volume is a sum, $\Gamma_{\text{tr}}+\Gamma_{\text{sl}}$. At moderate $\text{Re}_{\lambda}$ we have \cite{Falkovich2002}
\begin{align}&&
\Gamma_{\text{tr}}=\Gamma_{\text{ST}}\left(\frac{\eta}{2R}
\right)^{\alpha},\ \ \Gamma_{\text{sl}}=C4\pi R^2 F l\langle n\rangle^2 \left(\frac{\eta}{l}\right)^{\alpha}. \label{Skes}
\end{align}
Here $\Gamma_{\text{ST}}\sim \sqrt{8\pi/15} (2R)^3\langle n\rangle^2/\tau_{\eta}$ is the famous Saffman-Turner collision rate of finite-size tracers in incompressible turbulence \cite{Saffman1956} and $(\eta/2R)^{\alpha}$ is the preferential concentration factor due to centrifuge effect. The contribution of the sling effect $\Gamma_{\text{sl}}$ includes the order one constant $C$. That provides the volume fraction where the sling effects occurs as $C F\tau$. Here we observed that since duration of the sling event is of order $\tau$, then the fraction of time during which the particle is inside a sling is the frequency of sling events $F$ times $\tau$ times a numerical constant of order one $C$. We assumed that by ergodicity the fraction of time and the fraction of space where the sling event holds coincide.
Finally, $l$ in Eq.~(\ref{Skes}) is the effective spatial extension of the vortices that create sling effect. The turnover time of these vortices is of order $\tau$, which using the standard phenomenology of turbulence in the sense of Kolmogorov 41 \cite{Frisch1995} would give 
$l={\tilde C}\eta \text{St}^{1/2}$ where ${\tilde C}$ is a dimensionless constant of order one, see, e.g., \cite{Frisch1995,Falkovich2002}. This estimate probably must be modified in the limit of large Reynolds numbers where the non-trivial geometry of the vortices and intermittency would produce 
$l={\tilde C}\eta \text{St}^{1/2} \text{Re}_\lambda^{-\Delta}$ with $\Delta\neq 0$. 

In the flow region the particles move according to a prescribed compressible flow and, as it is generally true in such situations, they localize on a random strange attractor which is multifractal. The exponent $\alpha$ in Eq.~(\ref{Skes}) is the codimension of this multifractal, given by the space dimension $3$ minus the so-called correlation dimension of the multifractal. In the limit of small $\text{St}$ the particles' flow is weakly compressible and a universal formula holds \cite{Fouxon2012}  
\begin{eqnarray}&&
\alpha=\frac{2\sum_{i=1}^3\lambda_i}{\lambda_3}, \label{Sals}
\end{eqnarray}
where $\lambda_i$ are the Lyapunov exponents of the particles' flow. Here $\lambda_3$ provides the logarithmic divergence rate of close particles backwards in time and $\sum_{i=1}^3\lambda_i$ gives the logarithmic growth rate of small volumes of particles. It is readily seen that the right-hand side of Eq.~(\ref{Sals}) is the compressibility ratio of characteristic values of the flow divergence and gradients. It is useful because it seems to hold quite well also when the flow is not so weakly compressible \cite{Fouxon2022}. When $\text{Re}_{\lambda}$ is moderate an explicit formula holds \cite{Falkovich2002}  
\begin{align}\alpha=\frac{\tau^2}{\lambda_3}\int_{-\infty}^{\infty} \left\langle \nabla_iu_k\nabla_ku_i(0)\nabla_iu_k\nabla_ku_i(t)\right\rangle dt\,.\label{SUr}
\end{align}
However, $\alpha$ is generally bounded between zero and three (it is positive so that the concentration field is fractal and correlations are singular at zero distance. It is smaller than three so that the integral of the pair-correlation function of concentration over the distance converges, as necessary due to finite total mass). The boundedness of $\alpha$ implies quite certainly that there exists a finite limit \cite{Fouxon2022}: $
\alpha^*(\text{St})=\lim_{\text{Re}_{\lambda}\to\infty} \alpha(\text{St}, \text{Re}_{\lambda})$. While \cite{Falkovich2002} observe a clear Reynolds number trend for small Reynolds numbers \cite{ireland2016} seem to indiciate saturation of the above limit at moderate $\text{Re}_{\lambda}$.  However, caution is needed since very slow dependencies of $\alpha$ on $\text{Re}_{\lambda}$ could occur (cf. with the statistics of turbulent spectrum in the viscous range \cite{Ishihara2007}).

\begin{figure}
    \centering
    \includegraphics[width=\textwidth]{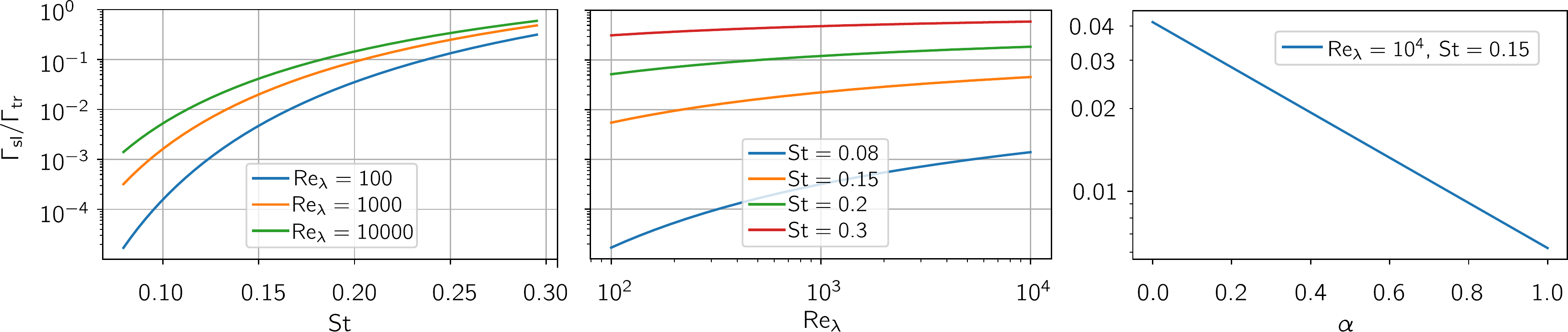}
    \caption{The ratio between sling and tracer contribution Eq.~\eqref{Sratios} for $\alpha=0$ once as a function of the Stokes number (left) and once as a function of Reynolds number (center). In general, the relevance of sling events decreases for smaller Stokes numbers. However, for smaller Stokes numbers the relevance increases stronger with Reynolds. (Right) We show for one selected Reynolds and Stokes number the influence of different values of $\alpha$, which can almost lead to an order of magnitude difference.}
    \label{fig:S5}
\end{figure}

Due to the boundedness of $\alpha$, self-consistency demands that the characteristic value of $|\nabla \bm u|$ that determines the correlation function is much smaller than $1/\tau$. This condition breaks down at moderate $\text{Re}_{\lambda}$ as discussed in \cite{Fouxon2022}:
we observe that $\alpha$ is determined by a high-order moment of velocity gradients. Therefore, due to intermittency, the characteristic value of velocity gradients that determine the average in Eq.~(\ref{SUr}) grows with $\text{Re}_{\lambda}$. The average would give at high Reynolds numbers $\alpha\propto \text{St}^2 \text{Re}_{\lambda}^{\delta}$ where $\delta>0$. Thus, Eq.~(\ref{SUr}) gets inconsistent as higher $\text{Re}_{\lambda}$ are considered as long as $\text{St}$ is finite. For however large but fixed $\text{Re}_{\lambda}$ though, Eq.~(\ref{SUr}) holds true provided that we take an appropriately small $\text{St}$, see \cite{Fouxon2022}. For $\text{St}\sim0.1$, it was checked to agree with the numerical value at $\text{Re}_{\lambda}=21$ within $4$ per cent, see \cite{Falkovich2004}. However, if we consider $\text{Re}_{\lambda}=47$  about 45 per cent deviations of measured $\alpha$ from Eq.~\eqref{SUr} were observed.

In the following, we introduce the critical Reynolds number $\text{Re}_{\text{cr}}(\text{St})$ defined by the condition that the correlation function in Eq.~(\ref{SUr}) is defined by vortices with characteristic turnover time $\tau$ at $\text{Re}_{\lambda}\sim \text{Re}_{\text{cr}}(\text{St})$. 
So the results of \cite{Falkovich2004} indicate that $\text{Re}_{\text{cr}}(\text{St}\sim 0.1)\sim 50$. 
Coincidentally, velocity gradients of order $1/\tau$ and larger are precisely those which lead to sling events. 
 Yet the clustering  in the smooth flow region of space is still determined by the boundary of those sling regions. That boundary consists of vortices with turnover time of order $\tau$ and smaller. If clustering is dominated by those it implies that the majority of close by droplets within the smooth flow region resides precisely in those boundaries. 
 Accordingly, also $\Gamma_{\text{tr}}$ is determined by those.
 Thus, for $\text{St}\sim 0.1$ we find from the above considerations that at $\text{Re}_{\lambda}\gtrsim 10^2$ all collisions are due to resonant vortices with characteristic time $\tau$, i.e., the collisions in slower vortices give negligible contribution to $\Gamma$. The characteristic value of velocity gradient determining $\Gamma_{\text{tr}}$ is $1/\tau$. In $\Gamma_{\text{ST}}$ the characteristic value of the velocity gradient is estimated by $1/\tau_\eta$. This increases this component of the collision kernel compared to the Saffman Turner kernel by a factor of order $\text{St}^{-1}$ giving,
\begin{align}
\Gamma_{\text{tr}}=\frac{C' \langle n\rangle^2}{\text{St}}\Gamma_{\text{ST}}
\left(\frac{\eta}{2R}
\right)^{\alpha},\ \ \text{Re}_\lambda\gg \text{Re}_{\text{cr}}(\text{St}), \label{Sgtr}
\end{align}
where $C'$ is a constant of order one. We observe from the definition that $\text{Re}_{\text{cr}}(St=0)=\infty$. It is probable that the divergence obeys asymptotic power-law dependence $\text{Re}_{\text{cr}}(\text{St})\sim \text{St}^{-\kappa}$ with a constant positive $\kappa$, however this will not be pursued here. 

We find from Eqs.~(\ref{Skes}) and (\ref{Sgtr}) that the ratio of contributions of particles that move in the multi- and single- valued regions of the particles' flow to the collision rate reads 
\begin{align}
\frac{\Gamma_{\text{sl}}}{\Gamma_{\text{tr}}}
=
\frac{C''4\pi R^2 F l \text{St}}{\Gamma_{\text{ST}}} \left(\frac{2R}{l}\right)^{\alpha}, \label{Sratios}
\end{align}
where $C''=C/C'$ is the ratio of two order one constants. The formulas for $\Gamma_{\text{sl}}$ and $\Gamma_{\text{tr}}$ contain unknown factors that could change them by order of magnitude. These are the constant $C''$ and exponents $\Delta$ and $\alpha$. Quite small differences of the latter exponents could change the rate strongly, as shown at the example of $\alpha$ in Fig.~\ref{fig:S5}. This demonstrates how far we still are from having a good idea of the magnitude of the effect of turbulence on the rain formation (and we have not started to speak yet of the hydrodynamic interactions that are disregarded in the study of geometric collisions).

Considering the difficulties described above, here we make the simplest possible estimates, in order to get at least some guiding in an unknown situation. In the following, we assume $\Delta=0$ i.e. using the phenomenology of turbulence in the sense of Kolmogorov 1941 \cite{Frisch1995}, thereby, neglecting potential intermittency effects. In Fig.~\ref{fig:S5}, we show the dependence of the ratio between the sling and tracer contribution, Eq.~\eqref{Sratios}, on Reynolds and Stokes number. Here, the value for $\alpha$ is set to zero. Generally we observe that the relevance of sling events increases with the Stokes number. The Reynolds dependence is however more pronounced for small Stokes numbers i.e.~the ratio increases by an order of magnitude for $\text{St}=0.15$ between $\text{Re}_\lambda=10^2$ and $\text{Re}_\lambda=10^4$. This means that especially for small Stokes numbers the Reynolds numbers present in clouds have a large effect.

\end{document}